\documentclass[aps,prb,twocolumn,superscriptaddress]{revtex4-1}
\usepackage{hyperref}
\usepackage{dsfont}
\usepackage{amsmath}
\usepackage{amssymb}
\usepackage{graphicx}
\usepackage{color}

\newcommand{\diff}{\mathrm{d}}  
\newcommand{\e}{\mathrm{e}} 
\newcommand{\mbf}[1]{\mathbf{#1}}  
\newcommand{\sign}{\operatorname{sign}}
\newcommand{\Tr}{\operatorname{Tr}}
\renewcommand{\Im}{\operatorname{Im}}
\newcommand{\B}{\operatorname{B}} 

\newcommand{\up}{\uparrow}
\newcommand{\down}{\downarrow}
\newcommand{\Ham}{\mathcal{H}} 
\newcommand{\Ej}{E_{\mathrm{J}}} 

\newcommand{\qp}[1]{q^{\prime}_{#1}}  

\begin{document}


\title{Decay of plasmonic waves in Josephson junction chains}


\author{M. Bard}
\affiliation{Institut f\"ur Nanotechnologie, Karlsruhe Institute of Technology, 76021 Karlsruhe, Germany}
\author{I.~V. Protopopov}
\affiliation{Department of Theoretical Physics, University of Geneva, 1211 Geneva, Switzerland  }
\affiliation{Landau Institute for Theoretical Physics, 119334 Moscow, Russia}
\author{A.~D. Mirlin}
\affiliation{Institut f\"ur Nanotechnologie, Karlsruhe Institute of Technology, 76021 Karlsruhe, Germany}
\affiliation{Institut f\"ur Theorie der Kondensierten Materie, Karlsruhe Institute of Technology, 76128 Karlsruhe, Germany} 
\affiliation{Petersburg Nuclear Physics Institute, 188350 St. Petersburg, Russia}
\affiliation{Landau Institute for Theoretical Physics, 119334 Moscow, Russia}


\date{\today}

\begin{abstract}
We study the damping of plasma waves in linear Josephson junction chains as well as in two capacitively coupled chains. In the parameter regime where the ground capacitance can be neglected, the theory of the antisymmetric mode in the double chain can be mapped onto the theory of a single chain. We consider two sources of relaxation: the scattering from quantum phase slips (QPS) and the interaction among plasmons related to the nonlinearity of the Josephson potential. The contribution to the relaxation rate $1/\tau$ from the nonlinearity scales with the fourth power of frequency $\omega$, while the phase-slip contribution behaves as a power law with a non-universal exponent. In the parameter regime where the charging energy related to the junction capacitance is much smaller than the Josephson energy, the amplitude of QPS is strongly suppressed. This makes the relaxation mechanism related to QPS efficient only at very low frequencies. As a result, for chains that are in the infrared limit on the insulating side of the superconductor-insulator transition,  the quality factor $\omega\tau$ shows a strongly non-monotonic dependence on frequency, as was observed in a recent experiment.

\end{abstract}


\maketitle


\section{Introduction}
Josephson-junction (JJ) chains constitute an ideal playground to study a wealth of fascinating physical effects. Parameters of these systems can be engineered in a controllable way, leading to the emergence of various physical regimes. In chains with the charging energy dominating over the Josephson energy, the Coulomb blockade is observed\cite{HavilandDelsing96} and a thermally activated conductance is found\cite{ZimmerEtAl13} at low bias. Moreover, the critical voltage at which the  conduction sets in, is governed by the depinning physics \cite{VogtEtAl15,CedergenEtAl17}. In the opposite limit, where the Josephson energy is the dominant energy scale, superconducting behavior in the current-voltage characteristics is observed \cite{ChowEtAl98,ErguelEtAl13b}. Deep in the superconducting regime, plasmonic waves (small collective  oscillations of the  superconducting phase) are well-defined excitations above the classical superconducting ground state.
The non-perturbative processes in which the phase difference across one of the junctions changes by $2\pi$---the  so called quantum phase slips\cite{BradleyDoniach84,MatveevEtAl02,PopEtAl10,RastelliEtAl13,ErguelEtAl13} (QPS)---are exponentially rare. 
Upon lowering the Josephson energy, QPS proliferate and eventually lead to the  superconductor-insulator transition \cite{BradleyDoniach84,ChoiEtAl98,ChowEtAl98,HavilandEtAl00,KuoChen01,MiyazakiEtAl02,TakahideEtAl06,RastelliEtAl13,BardEtAl17,CedergenEtAl17} (SIT) that occurs when the charging and Josephson energies are of the same order. 

Disorder plays an important role in JJ chains.  The effect of disorder was discussed in the context of the persistent current in closed JJ  rings in Ref.~\onlinecite{MatveevEtAl02}. More recently, the impact of various types of disorder on the SIT was studied in  Ref.~\onlinecite{BardEtAl17}. Remarkably, the most common type of disorder, random off-set charges, works to {\it enhance} superconducting correlations. The mechanism behind this effect is the loss of coherence of QPS due to a disorder-induced random phase, see also Ref.~\onlinecite{SvetogorovBasko18}. 



In recent years, properties of JJ chains under microwave irradiation have attracted considerable interest. Microwave radiation leads to quantized current steps in the current-voltage characteristics that were argued to be promising for metrological applications \cite{GuichardHekking10}. Another interesting direction in this context is the field of circuit quantum electrodynamics where novel regimes can be reached \cite{ZhangEtAl14,MartinezEtAl18}. JJ chains can be further employed to provide a tunable ohmic environment \cite{RastelliPop18}. This environment is realized by two parallel chains that are coupled capacitively to each other, and inductively to transmission lines. 

A similar setup was  used in Ref.~\onlinecite{KuzminEtAl18} to probe the reflection coefficient of a JJ double chain  under microwave irradiation. Two parallel chains are  short-circuited at one end while being coupled at the other end to a dipole antenna that can excite antisymmetric plasma waves (i.e., those with opposite amplitudes in the two chains). The whole sample is placed in a metallic waveguide which reduces the influence of external disturbances. Resonances corresponding to individual plasmonic modes at quantized momenta are clearly observed. This enables the reconstruction of  the energy spectrum of the plasma waves. Because of finite damping, the resonances in the reflection coefficient acquire a finite width. By measuring the modulus and the phase of the reflection coefficient, the internal damping could be disentangled from the external losses such as the leakiness of the waveguide or the damping of the transmission line. For chains with a large Josephson energy, the experimentally found quality factor (inverse linewidth multiplied by mode frequency) increased with lowering frequency of the microwave radiation. When the Josephson energy was reduced, the curves became flat and eventually showed a tendency to drop at low frequencies. This behavior was interpreted in Ref.~\onlinecite{KuzminEtAl18} as a signature of the SIT.
It was noted, however, that, in contrast to theoretical predictions, the observed behavior is controlled 
by the short-wavelength rather than the long-wavelength part of the Coulomb interaction in the chain. In particular, the ``superconducting'' behavior with the quality factor growing at low frequencies was observed in the range of parameters corresponding to the insulating phase of the chain.  


The purpose of this work is to provide  theoretical understanding of the effects related to the internal damping of plasma waves in JJ chains. We study two models: (i) a single linear chain, and (ii)  a double chain of JJs, as in the experiment of Ref.~\onlinecite{KuzminEtAl18}. It is shown that the effective theory for the antisymmetric mode of the double chain can be mapped onto a theory for a single chain if the capacitance to the ground can be neglected.
We identify two sources  that lead to the decay of plasmons: (i) the scattering of plasmons induced by QPS and (ii) the interaction of plasmon modes via ``gradient'' anharmonicities. We find the contribution to the relaxation rate of a plasma wave for both kinds of damping mechanisms. 
The ``gradient'' nonlinearities are always irrelevant in the renormalization-group sense and the corresponding relaxation rate vanishes as $\omega^4$ at low frequencies. From the SIT point of view, this behavior can be viewed as  ``superconducting''. 
On the other hand, the contribution of QPS processes  can show both ``superconducting'' and ``insulating'' trends depending on the parameters of the model. 
The QPS contribution  is, however, strongly suppressed if the Josephson energy is much larger than the charging energy associated with the junction capacitance that controls the short-wavelength part of the Coulomb interaction. The combination of the two mechanisms (QPS and ``gradient'' anharmonicities) can thus lead to a  change of the trend from ``insulating'' to ``superconducting'' at intermediate frequencies. This mimics a SIT in the intermediate frequency range, although the system is in fact deeply in the insulating phase from the point of view of its infrared behavior. 

The paper is structured as follows. In Sec.~\ref{Sec:Model} we introduce lattice models for a single JJ chain and for two capacitively coupled chains, and derive the effective low-energy field theory. Sec.~\ref{Sec:Relaxation} discusses two mechanisms contributing to the finite lifetime of the plasmonic waves in JJ chains. The scattering off QPS is studied in Sec~\ref{Sec:phase_slips}, and the decay  because of interactions between plasmonic waves is analyzed in Sec.~\ref{Sec:nonlinearities}. We analyze the interplay of both mechanisms in Sec.~\ref{Sec:interplay}. In Sec.~\ref{Sec:Summary} we summarize the main results of the paper and compare them to experimental findings. Technical details can be found in the appendix.



\section{Lattice models and low-energy theory}
\label{Sec:Model}

\begin{figure}
\centering
\includegraphics[scale=0.3]{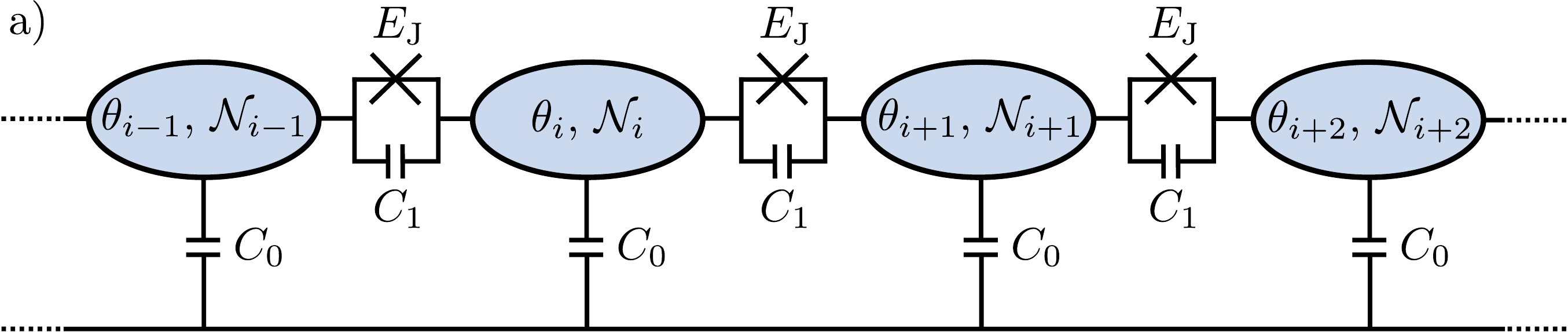}
\\
\vspace{0.8cm}
\includegraphics[scale=0.3]{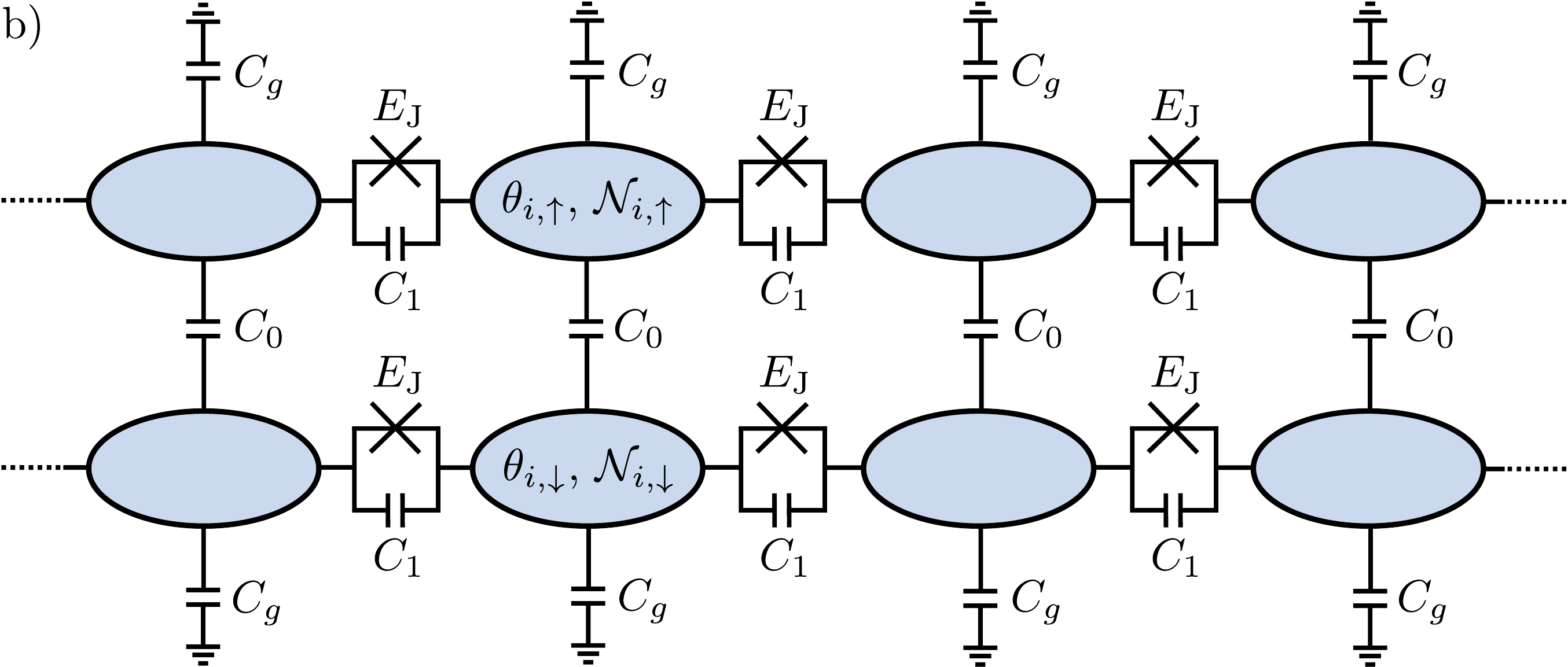}
\caption{Schematic representation of the two devices under consideration: a) A single chain of Josephson junctions. The capacitance to the ground is denoted by $C_0$, and the junction capacitance by $C_1$. The canonically conjugated variables are the superconducting phase $\theta_i$ and the number of Cooper pairs $\mathcal{N}_i$ of the $i$-th island. b) Two capacitively coupled chains. Here, the capacitance to the ground is denoted by $C_g$ and the interchain capacitance by $C_0$. The additional index $\up,\down$ discriminates between the variables of the two chains. 
\label{Fig:schematic-system}}
\end{figure}

In this work, we consider two closely related systems: a single linear chain of Josephson junctions depicted in Fig.~\ref{Fig:schematic-system}a and a device consisting of two capacitively coupled chains shown in Fig.~\ref{Fig:schematic-system}b. 
We are interested in their 
effective low-energy description. For a single chain of Josephson junctions with Coulomb interaction and disorder, Fig.~\ref{Fig:schematic-system}a, the field theory was constructed previously in Ref.~\onlinecite{BardEtAl17}. We  briefly recall this derivation below and extend the theory by including the terms accounting for gradient nonlinearities.  We then show that, up to numerical coefficients, the same effective description applies to 
the antisymmetric  mode of the double JJ chain of 
Fig.~\ref{Fig:schematic-system}b, provided that $C_g\ll C_0$.  

In the case of a single chain, we denote   by $C_1$ and $C_0$ the junction capacitance and the capacitance to the ground, respectively. Tunnel barriers between the islands allow for hopping of Cooper pairs along the chain. The number of Cooper pairs $\mathcal{N}_{i}$ and the superconducting phase $\theta_i$ of the $i$-th island satisfy the canonical commutation relation, $[\mathcal{N}_i,\theta_{j}]=i\delta_{i,j}$. Besides the Josephson energy $\Ej$ that quantifies the hopping strength of Cooper pairs, there are the two charging energy scales $E_0=(2e)^2/C_0$ and $E_1=(2e)^2/C_1$, where $e$ denotes the elementary charge. 
The charging energy $E_1$ quantifies the strength of the Coulomb interaction at short scales, while the energy $E_0$ controls the  Coulomb-interaction strength  in the infrared and, in particular,  determines the position of the SIT\cite{ChoiEtAl98,BardEtAl17}.
The lattice Hamiltonian for this system has the form
\begin{equation}
\Ham=\frac{E_1}{2}\sum_{i,j}s_{i,j}^{-1}\mathcal{N}_{i}\mathcal{N}_j+\Ej\sum_i [1-\cos(\theta_{i+1}-\theta_i)],
\label{Eq:lattice-Hamiltonian-single-chain}
\end{equation}
where 
\begin{equation}
s_{i,j}=\frac{C_{i,j}}{C_1}=\left(2+\frac{1}{\Lambda^2}\right)\delta_{i,j}-\delta_{i,j+1}-\delta_{i,j-1}
\end{equation}
is the dimensionless capacitance matrix and $\Lambda=\sqrt{C_1/C_0}$ is the screening length for the 1D Coulomb interaction. 

In the low-energy limit, it is legitimate to replace the lattice Hamiltonian (\ref{Eq:lattice-Hamiltonian-single-chain}) by an effective continuum model. The latter is conveniently written\cite{BardEtAl17} in terms of the field $\phi(x)$ related to the density of Cooper pairs  ${\cal N}(x)$ by $\partial_x\phi(x)=-\pi {\cal N}(x)$.
 The action of the model reads\cite{BardEtAl17}
\begin{equation}
S=S_0+S_{\mathrm{ps}}.
\label{Eq:S_clean}
\end{equation}
The quadratic part of the action (in the imaginary-time representation, with temperature $T$ and Matsubara frequencies $\omega_n$)
\begin{equation}
S_0=\frac{1}{2\pi^2 u_0 K_0} T \sum_{\omega_n}\int\frac{\diff q}{2\pi}\left[\omega_n^2+\epsilon^2(q)\right]|\phi(q,\omega_n)|^2
\label{Eq:S_0}
\end{equation}
describes the plasma waves with the energy spectrum
\begin{equation}
\epsilon(q)=\frac{\omega_p|q|}{\sqrt{q^2+\alpha/\Lambda^2}},  \quad \omega_p=\sqrt{\Ej E_1}.
\label{Eq:dispersion}
\end{equation}
To facilitate our future discussion of the effective theory for the double chain setup of Fig.~\ref{Fig:schematic-system}b, we have introduced here a numerical  coefficient $\alpha$; in the present case of a single chain we have $\alpha\equiv 1$. The parameters $u_0$ and $K_0$ in Eq.~(\ref{Eq:S_0}) are given by  
\begin{equation}
u_0=\sqrt{\frac{\Ej E_0}{\alpha}},\qquad  K_0=\sqrt{\frac{\Ej}{\alpha\, E_0}}.
\label{Eq:parameters}
\end{equation}
Here $u_0$  is the velocity of low-energy plasmons with momentum $q\ll 1/\Lambda$ and $K_0$ is  the corresponding  Luttinger constant. 
Note that we measure all distances in units of the lattice spacing and set $\hbar = 1$, so that the velocity has the dimension of energy.


The second ingredient in Eq.~(\ref{Eq:S_clean}), $S_{\rm ps}$, describes QPS. In the absence of disorder it is given by\cite{BardEtAl17}
\begin{equation}
S_{\mathrm{ps}}=yu_0\int \diff x\diff\tau \cos\left[2\phi(x,\tau)\right],
\label{Eq:S_ps}
\end{equation}
where $\tau$ is the imaginary time and $y$ is the (ultraviolet) value of the QPS amplitude that is usually called ``fugacity''. This terminology is related to the fact that QPS can be considered as 
 vortices in the Euclidean version of the  $1+1$-dimensional quantum theory.  The fugacity $y$ for phase slips is exponentially small in the regime $\Ej \gg \min(E_1,E_0)$ where the superconducting correlations are (at least locally) well developed, 
\begin{equation}
y\propto \e^{-\zeta K}\,, \qquad K=\sqrt{\frac{\Ej}{\alpha\, E_0}+\frac{\Ej}{\alpha^2\, E_1}} \,.
\label{Eq:y}
\end{equation}
Here $K$ plays the role of the Luttinger constant for the ultraviolet plasmons (with $q\sim 1$), and  $\zeta$ is a numerical factor that depends on the screening length $\Lambda$ and  also on details of the ultraviolet cutoff scheme. Estimates for $\zeta$ in several limiting cases can be found in Refs.\onlinecite{BradleyDoniach84,ChoiEtAl98,MatveevEtAl02,RastelliEtAl13,SvetogorovBasko18}. 

Among various types of disorder that are present in experimental realizations of the JJ chains, the strongest and the most important one is the random stray charges.
Random stray charges $Q_i$ modify the kinetic energy term in the lattice Hamiltonian, Eq.~(\ref{Eq:lattice-Hamiltonian-single-chain}), according to 
\begin{equation}
\sum_{i,j}s_{i,j}^{-1}\mathcal{N}_{i}\mathcal{N}_j \ \ \longrightarrow\ \  \sum_{i,j}s_{i,j}^{-1}\left(\mathcal{N}_{i}-Q_i\right)\left(\mathcal{N}_j-Q_j\right).
\end{equation}
The wave function of the system accumulates  then an extra phase in the course of a QPS  due to the Aharonov-Casher effect\cite{AharonovCasher1984, MatveevEtAl02, Pop2012}. Accordingly, the QPS action in the effective theory  in the presence of charge disorder takes the  form 
\begin{equation}
S_{\mathrm{ps}}=yu_0 \int \diff x \diff \tau \cos\left[2\phi(x,\tau)-\mathcal{Q}(x)\right],
\label{Eq:S_ps_disordered}
\end{equation}
where
\begin{equation}
\mathcal{Q}(x)=2\pi \int_{-\infty}^{x} \diff x^{\prime} \, Q(x^{\prime}).
\end{equation}
For simplicity we assume Gaussian white noise disorder,
\begin{equation}
\left<Q(x)\right>_Q=0, \quad \left<Q(x)Q(x^{\prime})\right>_Q=\frac{D_Q}{2\pi^2} \delta(x-x^{\prime}).
\end{equation}

In the Hamiltonian language, the action \eqref{Eq:S_clean} corresponds to
\begin{equation}
\Ham=\Ham_0+\Ham_{\mathrm{ps}},
\label{Eq:Hamiltonian}
\end{equation}
where the quadratic part is of the form
\begin{equation}
\Ham_0=\frac{1}{2\pi^2}\int\frac{\diff q}{2\pi}\left[\frac{\epsilon^2(q)}{u_0K_0}|\phi(q)|^2+u_0 K_0 q^2 |\pi\theta(q)|^2\right],
\end{equation}
and the QPS contribution reads
\begin{equation}
\Ham_{\mathrm{ps}}=yu_0\int \diff x \cos[2\phi(x)-\mathcal{Q}(x)].
\label{Eq:Hamiltonian_ps}
\end{equation}

Equations (\ref{Eq:S_clean}), (\ref{Eq:S_0}) and (\ref{Eq:S_ps_disordered}) [the latter one reduces to Eq.~(\ref{Eq:S_ps}) in the clean case] constitute the low-energy description of a JJ chain as derived in Ref.~\onlinecite{BardEtAl17}. Several remarks are in order here. First, 
in this work we will be interested in the physics at moderate wavelengths  $q\lesssim 1/\Lambda$ and will approximate the dispersion relation (\ref{Eq:dispersion}) by its expansion at small $q$:
\begin{eqnarray}
\epsilon(q)&\approx& u_0(1-q^2 l^2)|q|,
\label{Eq:dispersion1}
\\
 l&=&\Lambda/\sqrt{2\alpha}.
 \label{Eq:l}
\end{eqnarray} 
Here, the length $l$ (that differs from $\Lambda$ only by a numerical coefficient) sets the scale for the bending of the plasmonic dispersion relation. 

Second, the only nonlinearity in our effective action at this stage is due to the QPS. 
In the ultimate infrared limit  $q\ll 1/\Lambda$ (or, in fact, for all $q$ in the case of short-range charge-charge interaction, $\Lambda\lesssim 1$) the effective action Eqs.~(\ref{Eq:S_clean}), (\ref{Eq:S_0}) and (\ref{Eq:S_ps}) reduces to the standard sine-Gordon theory and describes the superconductor-insulator transition (SIT) that occurs \cite{ChoiEtAl98} at $K_0=2/\pi$. 
The SIT is driven by the QPS and in this sense they constitute the most important anharmonicity in the system. Other non-harmonic terms are, however, also possible. For example, the expansion of the Josephson coupling in Eq.~(\ref{Eq:lattice-Hamiltonian-single-chain}) to the next-to-leading order provides the following contribution to the effective Hamiltonian: 
\begin{equation}
\Ham_{\mathrm{nl}}=-\frac{\Ej}{\alpha^3\, 4!}\int \diff x \left(\partial_x\theta\right)^4,
\label{Eq:nonlinearity}
\end{equation}
which is associated with the action
\begin{equation}
S_{\mathrm{nl}}=-\frac{\alpha}{4!\pi^4 \Ej^3}\int \diff x \diff \tau \left(\partial_{\tau}\phi\right)^4.
\label{Eq:nonlinearity-action}
\end{equation}

In contrast to the QPS term (\ref{Eq:Hamiltonian_ps}), the nonlinearity (\ref{Eq:nonlinearity}) and all other non-linear terms that can be added to the effective Hamiltonian are built out of the local charge 
${\cal N}\propto \partial_x\phi$ and current ${\cal J}\propto \partial_x\theta$ densities. Therefore, they always contain high powers of gradients and are irrelevant in the renormalization-group sense. We will occasionally refer to the anharmonicities of these types as to ``gradient''  anharmonicities to distinguish them from the QPS. The gradient anharmonicities are always unimportant at lowest energies. Yet, as  we will see in Sec.~\ref{Sec:nonlinearities}, they can  control the decay of plasmons at sufficiently high frequencies if the bare value of the QPS amplitude $y$ is small.

We thus conclude that the effective action  describing a chain of Josephson junctions is given by
\begin{equation}
S=S_0+S_{\rm ps}+S_{\rm nl}.
\label{Eq:SFinal}
\end{equation}
Here, $S_0$ and $S_{\rm ps}$ are given by Eqs.~(\ref{Eq:S_0}) and (\ref{Eq:S_ps_disordered}), respectively. As for the ``gradient''  anharmonicities represented by $S_{\rm nl}$, we will use Eq.~(\ref{Eq:nonlinearity-action}) as a specific form. We argue in Sec.~\ref{Sec:nonlinearities} that our main conclusions are insensitive to this particular choice. 

Let us now turn to the discussion of the double-chain setup of 
Fig.~\ref{Fig:schematic-system}b.
In this case  we denote by $C_g$ and $C_0$ 
the capacitance to the ground and the interchain capacitance, respectively.
 The lattice Hamiltonian for the double chain reads
\begin{equation}
\begin{split}
\Ham=\frac{E_1}{2}\sum_{i,j}\sum_{\sigma,\sigma^{\prime}=\up,\down}&[\mathcal{S}^{-1}]_{\sigma,\sigma^{\prime}}(i,j)\mathcal{N}_{i,\sigma}\mathcal{N}_{j,\sigma^{\prime}}
\\
&+\Ej \sum_{i,\sigma}[1-\cos(\theta_{i+1,\sigma}-\theta_{i,\sigma})] ,
\end{split}
\label{Eq:lattice-Hamiltonian-double-chain}
\end{equation}
with
\begin{equation}
\mathcal{S}(i,j)=\tilde{s}_{i,j}
\begin{pmatrix}
1 & 0
\\
0 & 1
\end{pmatrix}
+\frac{C_0}{C_1} \delta_{i,j}
\begin{pmatrix}
1 & -1
\\
-1 & 1
\end{pmatrix}
\end{equation}
and
\begin{equation}
\tilde{s}_{i,j}=(2+C_g/C_1)\delta_{i,j}-\delta_{i,j+1}-\delta_{i,j-1}.
\end{equation}
Here, the indices $\up,\down$  refer to the two chains. 

In the Gaussian approximation the spectrum of the Hamiltonian (\ref{Eq:lattice-Hamiltonian-double-chain}) consists of two modes, symmetric and antisymmetric, that are analogous to the charge and spin modes in a spinful Luttinger liquid. In this work we are interested in the physics of the antisymmetric mode that can be excited in the system by coupling to a dipole antenna \cite{KuzminEtAl18}. To simplify the analysis, we assume further that  $C_g\ll C_0$.
It is not clear to us how well is this assumption satisfied in the experiments of Ref.~\onlinecite{KuzminEtAl18}; we believe it to be, however, of minor importance for our results. Specifically, our analysis should remain applicable, up to modifications in numerical coefficients of order unity, also for $C_g\sim C_0$.  The condition $C_g\lesssim C_0$ corresponds to sufficiently well coupled chains, with large splitting between the symmetric and antisymmetric modes\footnote{The opposite case, $C_0\ll C_g$, corresponds to nearly decoupled chains, in which case the physics is more naturally described in terms of modes corresponding to individual chains rather than in terms of symmetric and antisymmetric modes. In the limit $C_0=0$ the decoupling is complete, and the analysis for a single chain applies, with $C_g$ playing the role of $C_0$.}.  

  In full analogy to the spin-charge separation in quantum wires, the (low-momentum) velocity of the symmetric (``charge'') mode, $u_{\rm ch}=2\sqrt{e^2\Ej/C_g}$ greatly exceeds, under the assumption $C_g\ll C_0$, the velocity of the antisymmetric (``spin'') mode, $u_{\rm s}=\sqrt{2e^2 \Ej/C_0}$. This observation allows one to integrate out the charge mode and formulate the effective description of the system in terms of the antisymmetric mode alone. Details of this derivation are presented in appendix~\ref{App:Field_theory}. It turns out that, just as in the case of a single JJ chain, the effective theory is given by Eqs.~(\ref{Eq:SFinal}), (\ref{Eq:S_0}), (\ref{Eq:S_ps_disordered}) and (\ref{Eq:nonlinearity-action}), with the parameters given by Eqs.~(\ref{Eq:dispersion1}),  (\ref{Eq:l}) and (\ref{Eq:parameters}).  The only difference is the value of the numerical factor $\alpha$ that should now be set to $2$.

Equations (\ref{Eq:SFinal}), (\ref{Eq:S_0}), (\ref{Eq:S_ps_disordered}) and (\ref{Eq:nonlinearity-action}) constitute the starting point for our analysis of the decay of plasmonic excitations in the setups of Fig.~\ref{Fig:schematic-system}. We carry out this analysis in Sec.~\ref{Sec:Relaxation}.

\section{Relaxation of plasmonic waves}
\label{Sec:Relaxation}

Plasmonic waves, which are long-wavelength excitations above the superconducting ground state, are subjected to interaction. As a result, once excited  by, e.g., a microwave, a plasma wave can decay into several plasmons of lower energy. The two anharmonic terms in the action (\ref{Eq:SFinal}) provide two 
mechanisms for the decay of plasmons: interaction with QPS and ``gradient'' anharmonicities. We analyze these channels of plasmon decay one by one in Secs.~\ref{Sec:phase_slips} and \ref{Sec:nonlinearities}, respectively.
The interplay of the two mechanisms is discussed in Sec.~\ref{Sec:interplay}.

\subsection{Relaxation due to phase slips}
\label{Sec:phase_slips}

We start with the discussion of the relaxation processes related to the scattering off QPS. 
Our analysis follows closely the one of Refs.~\onlinecite{Oshikawa2002, Rosenow2007}.
The curvature of the plasmonic spectrum, as quantified by the length $l$ in Eq.~(\ref{Eq:dispersion1}) is of minor importance here and for the purpose of this section we approximate the plasmonic spectrum 
by 
\begin{equation}
\epsilon(q)=u_0|q|.
\end{equation}
Correspondingly, the Gaussian action takes the form
\begin{equation}
S_0=\frac{1}{2\pi^2 u_0 K_0} \int \diff x \diff \tau \left[u_0^2(\partial_x\phi)^2+(\partial_{\tau}\phi)^2\right].
\label{Eq:S_0-Luttinger-liquid}
\end{equation}

 A formal expansion of the QPS action, Eq.~\eqref{Eq:S_ps}, in powers of $\phi$ shows that the plasmon can decay into an arbitrary large number of low-energy plasmons. We will determine directly the sum of all those contributions. This decay rate can be conveniently calculated from the imaginary part of the self-energy (of the Fourier transform) of the correlation function
\begin{equation}
G(\mathbf{r})=\left<\left<\phi(\mathbf{r})\phi(0)\right>_{S}\right>_{Q},
\label{Eq:Green-function}
\end{equation}  
where $\mathbf{r}=(x,\tau)$ and $\left<\cdot\right>_S$ denotes the average with respect to the full action, $S=S_0+S_{\mathrm{ps}}$. On the Gaussian level, the imaginary-time Green function reads, in Fourier space,
\begin{equation}
G_0(\mathbf{q})=\frac{\pi^2 u_0 K_0}{\omega_n^2 + u_0^2 q^2}, \qquad \mbf{q}=(q,\omega_n),
\end{equation}
where $\omega_n$ is the Matsubara frequency. 
With the help of the self-energy, the full Green function can be expressed as
\begin{equation}
G(\mbf{q})=\frac{1}{G_0^{-1}(\mbf{q})-\Sigma(\mbf{q})}=\frac{\pi^2 u_0 K_0}{\omega_n^2+u_0^2 q^2 - \pi^2 u_0 K_0 \Sigma(\mbf{q})}.
\end{equation}
The inverse lifetime of an excitation with energy $\omega$ is related to the imaginary part of the retarded self-energy on the mass shell:
\begin{equation}
\frac{1}{\tau(\omega)}=\frac{\pi^2 K_0 u_0}{2\omega}\Im\Sigma^R(q=\omega/u_0,\omega).
\label{Eq:inverse-lifetime}
\end{equation}
 In the  following, we calculate the self-energy perturbatively in the QPS fugacity $y$. The self-energy in the Matsubara space-time,  $\Sigma(\mbf{r})$, can be extracted from the perturbative expansion of the Green function, Eq.~\eqref{Eq:Green-function},
\begin{equation}
G(\mbf{r})=G_0(\mbf{r})+\int \diff^2 r_1 \diff^2 r_2 G_0(\mbf{r}-\mbf{r}_1)\Sigma(\mbf{r}_1-\mbf{r}_2)G_0(\mbf{r}_2),
\end{equation}
with the following result:
\begin{equation}
\begin{split}
\Sigma(\mbf{r})=2y^2 &u_0^2 \Bigl[ \e^{-2 C_0(\mbf{r})-D_Q|x|}
\\
&-\delta(\mbf{r}) \int \diff^2 r_0 \e^{-2C_0(r_0)-D_Q |x_0|} \Bigr]+\mathcal{O}(y^4).
\label{Eq:self-energy}
\end{split}
\end{equation}
Here, the exponential contains the correlation function
\begin{align}
C_0(\mbf{r})&=\frac{2}{\beta N_x}\sum_{\mbf{q}}(1-\cos \mbf{qr})G_0(\mbf{q})
\\
&=\frac{\pi K_0}{2}\ln\left[\frac{u_0^2 \beta^2}{\pi^2}\sinh\left(z_{+}\right)\sinh\left(z_{-}\right)\right]
\end{align}
with
\begin{equation}
z_{\pm}=\frac{\pi}{u_0\beta}(x\pm i u_0 \tau),
\end{equation}
$N_x\gg 1$ denotes the number of junctions, and $\beta$ is the inverse temperature. The result for the self-energy in the imaginary time $\tau$ should be analytically continued to real time $t$ and then Fourier-transformed. The $\tau$-dependence of the first term in Eq.~(\ref{Eq:self-energy}) is determined by the 
following Matsubara time-ordered correlation function
\begin{equation}
\chi^T(x,\tau)=\e^{-2C_0(x,\tau)}.
\end{equation}
The retarded version of this correlation function can be obtained in the standard way \cite{GiamarchiBook}. We find
\begin{equation}
\chi^R(x,t)=\frac{2\Theta(t)\Theta\left(u_0 t -|x|\right) \sin\left(\pi^2 K_0\right)\left(\frac{\pi}{\beta u_0}\right)^{2\pi K_0}}{\left|\sinh \frac{\pi}{u_0\beta}\left(x+u_0 t\right)\sinh \frac{\pi}{u_0\beta}\left(x- u_0 t\right)\right|^{\pi K_0}},
\end{equation}
where $\Theta$ denotes the Heaviside step function. In order to extract the lifetime, we need to know the imaginary part of the self-energy in Fourier space. The second term on the RHS of Eq.~\eqref{Eq:self-energy} does not contribute to the imaginary part of $\Sigma$. The imaginary part of the self-energy in Fourier space can thus be obtained via
\begin{equation}
\Im \Sigma^R(q,\omega)=2u_0^2 y^2 \Im\!\! \int \!\!\diff x \diff t\, \e^{-i(q x- \omega t)}\chi^R(x,t)\e^{-D_Q|x|}.
\end{equation}
It is convenient to switch to the light-cone variables $z_{\pm}=\pi(u_0 t \pm x)/(u_0 \beta)$:
\begin{equation}
\begin{split}
\Im &\Sigma^R(q,\omega)=2u_0 y^2  \sin(\pi^2 K_0)\left(\frac{\pi}{u_0 \beta}\right)^{2\pi K_0-2}
\\
&\times\Im \int_0^{\infty}\diff z_{+}\int_0^{\infty}\diff z_{-} \frac{\exp\{i\frac{\beta}{2\pi}(\omega-u_0 q)z_{+}\}}{\left(\sinh z_{+}\right)^{\pi K_0}}
\\
&\times \frac{\exp\{i\frac{\beta}{2\pi}(\omega+u_0 q)z_{-}\}}{\left(\sinh z_{-}\right)^{\pi K_0}}\,\e^{-\frac{D_Q u_0 \beta}{2\pi}|z_{+}-z_{-}|}.
\end{split}
\label{Eq:Im-Sigma}
\end{equation}

Equations (\ref{Eq:inverse-lifetime}) and (\ref{Eq:Im-Sigma}) give the decay rate of plasma waves due to QPS. They can be further simplified in various limiting cases that we analyze below.

 
\subsubsection{Clean case\label{Sec:Clean-case}}

In the regime $D_Q \ll \mathrm{min}(q,T/u_0)$ we can set $D_Q=0$, and the integrations decouple. Performing the integrations, we obtain
\begin{equation}
\begin{split}
\Im\Sigma^R(q,\omega)&=2u_0 y^2 \sin(\pi^2 K_0)\left(\frac{2\pi}{u_0 \beta}\right)^{2\pi K_0-2}
\\
&\times \Im\Biggl\{ \B\left(1-\pi K_0,\frac{\pi K_0}{2}-i\frac{\beta }{4\pi}(\omega+u_0 q)\right)
\\
& \times\B\left(1-\pi K_0,\frac{\pi K_0}{2}-i\frac{\beta}{4\pi}(\omega-u_0 q)\right) \Biggr\},
\end{split}
\end{equation}
where 
\begin{equation}
\B(x,y)=\frac{\Gamma(x)\Gamma(y)}{\Gamma(x+y)}
\end{equation}
is the Euler Beta function. Making use of Eq.~\eqref{Eq:inverse-lifetime}, we extract the relaxation rate,
\begin{equation}
\frac{1}{\tau(\omega)}\sim u_0 y^2
\begin{cases}
\left(\frac{2\pi T}{u_0}\right)^{2\pi K_0-3}, & \omega \ll T,
\\
\frac{T}{u_0}\left(\frac{2\pi\omega T}{u_0^2}\right)^{\pi K_0 -2}, & \omega \gg T.
\end{cases}
\label{Eq:scaling-rate-clean}
\end{equation}
\begin{figure}
\centering
\includegraphics[width=220pt]{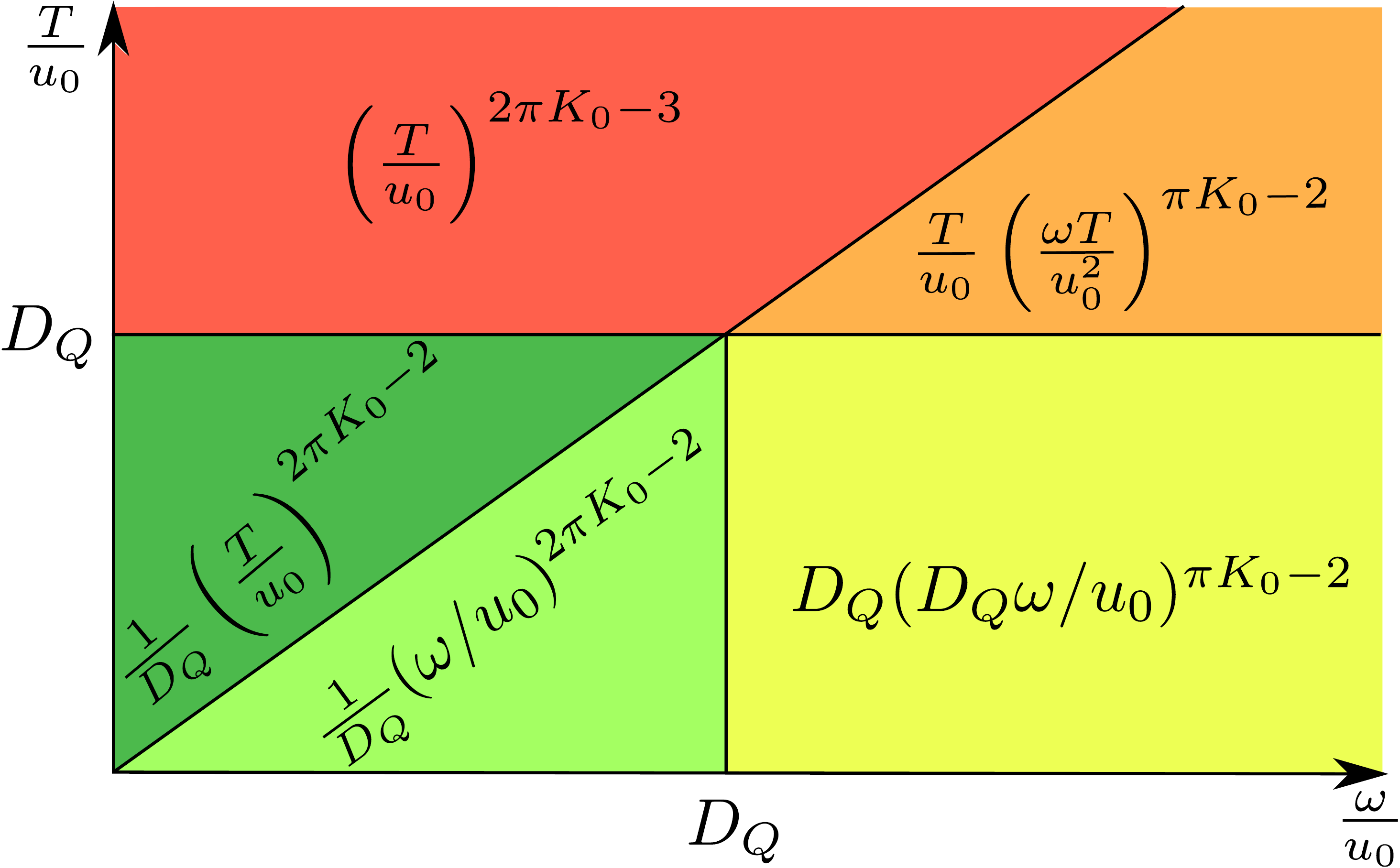}
\caption{Scaling behavior of the inverse relaxation time of plasmonic waves due the scattering from QPS in different regimes in the frequency-temperature plane. For each regime, the behavior of $1/(\tau u_0 y^2)$ is indicated.\label{Fig:rate-QPS-processes}}
\end{figure}

\subsubsection{Disordered case\label{Sec:Disordered-case}}
In the limit of strong disorder, $D_Q \gg \mathrm{max}(q,T/u_0)$, the main contribution of the integrations in Eq.~\eqref{Eq:Im-Sigma} comes from the region close to $z_{+}=z_{-}$. We find
\begin{equation}
\begin{split}
\Im \Sigma^{R}(q,\omega)\simeq 8 & u_0\frac{y^2}{D_Q}\sin(\pi^2 K_0) \left(\frac{2\pi}{u_0 \beta}\right)^{2\pi K_0 -1}
\\
& \times \Im \B\left(1-2\pi K_0,\pi K_0 -i \frac{\beta \omega}{2\pi}\right)
\end{split}
\end{equation}
for the imaginary part of the self-energy, which is independent of momentum. This leads to the following scaling of the relaxation rate  in the case of strong disorder, $D_Q \gg \mathrm{max}(q,T/u_0)$:
\begin{equation}
\frac{1}{\tau(\omega)}\sim u_0\frac{y^2}{D_Q}
\begin{cases}
\left(\frac{2\pi T}{u_0}\right)^{2\pi K_0-2}, &\omega \ll T,
\\
\left(\frac{\omega}{u_0}\right)^{2\pi K_0-2}, & \omega\gg T.
\end{cases}
\label{Eq:scaling-rate-strongly-disordered}
\end{equation}
For a moderate disorder strength, we need to consider two cases. If $q\ll D_Q \ll T/u_0$, the clean result given by the first line of Eq.~\eqref{Eq:scaling-rate-clean} remains valid. For $T/u_0\ll D_Q \ll q$ 
the integration over $z_{-}$ in Eq.~(\ref{Eq:Im-Sigma}) is cut at the upper limit by $\pi T/u_0 q \ll 1$. We can further neglect $z_{-}$ in the exponential function related to $D_Q$. This leads to the following behavior of the relaxation rate:
\begin{equation}
\frac{1}{\tau(\omega)}\sim u_0 y^2\ D_Q\left(\frac{D_Q\omega}{u_0}\right)^{\pi K_0 -2}, \quad T\ll u_0 D_Q \ll \omega.
\label{Eq:inter}
\end{equation}

Equations (\ref{Eq:scaling-rate-clean}), (\ref{Eq:scaling-rate-strongly-disordered}) and (\ref{Eq:inter}) 
give the relaxation rate of plasma waves due to QPS in different regimes and constitute the main result of this section. They are summarized in Fig.~\ref{Fig:rate-QPS-processes}.  The relaxation rate exhibits power-law scaling with  frequency, temperature and the disorder. The corresponding exponents are non-universal and are determined by the value of the Luttinger parameter~$K_0$. Deep in the superconducting phase of the JJ chain, $K_0\gg1$, the relaxation rate vanishes at low frequencies, while the opposite trend is predicted in the insulating phase with sufficiently small $K_0$.


\subsection{Relaxation due to ``gradient''  nonlinearities\label{Sec:nonlinearities}}

Let us now turn to the analysis of ``gradient'' anharmonicities described by the term $S_{\rm nl}$ in the effective action (\ref{Eq:SFinal}). They are irrelevant from the point of view of the renormalization group. However, at intermediate energy scales they contribute to the decay of the plasma waves on equal footing with QPS. We consider here the nonlinearity \eqref{Eq:nonlinearity} corresponding to the correction \eqref{Eq:nonlinearity-action} of the action, which arises as the quartic term of the expansion of the Josephson potential.

Perturbative treatment of the decay of plasmons caused by ``gradient'' anharmonicities was discussed in other contexts
in Refs.~\onlinecite{Apostolov2013, Lin2013, ProtopopovEtAl14}.  The perturbation theory turns out to be ill-defined in the case of a linear plasmonic spectrum and in this section we use the dispersion relation (\ref{Eq:dispersion1}) taking into account its finite curvature.

\begin{figure}
\centering
\includegraphics[scale=0.5]{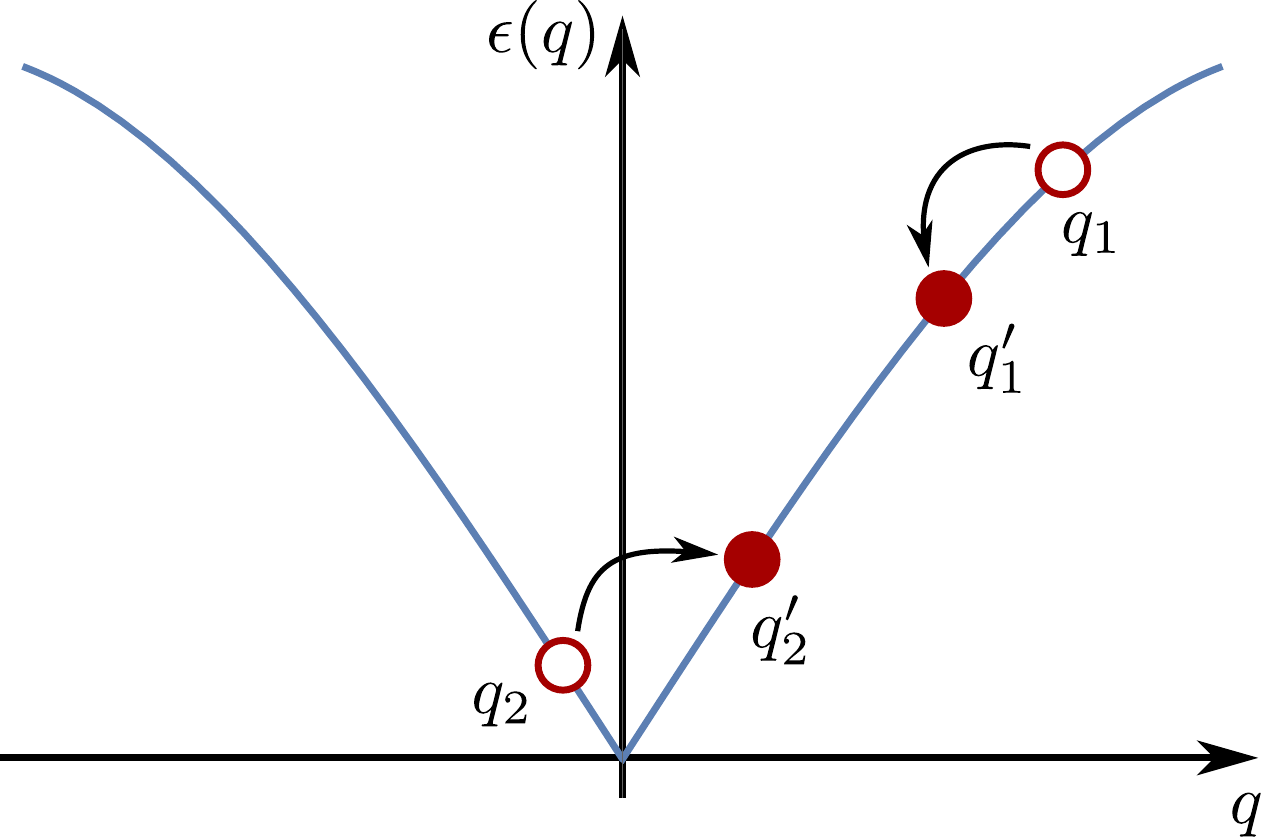}
\caption{Dominant relaxation process mediated by nonlinearities for a right moving plasmon with momentum $q_1$.\label{Fig:process}}
\end{figure}

In order to calculate the relevant matrix element for the relaxation process, it is convenient to express the superconducting phase $\theta$ through bosonic creation ($b_q^{\dagger}$) and annihilation operators ($b_q$), that obey the standard bosonic commutation relations. This decomposition is of the form\cite{GiamarchiBook}
\begin{equation}
\theta(x)=i\sqrt{\frac{\pi}{2N_x}}\sum_{q\neq 0}\frac{\sign(q)}{\sqrt{|q|}}\e^{-a |q|/2}\e^{i q x} (b_q^{\dagger}-b_{-q}),
\end{equation}
where $a$ is the ultraviolet cutoff that can be send to zero in this calculation, and $N_x$ is the number of junctions per chain. Our analysis below largely follows the approach of Ref.~\onlinecite{ProtopopovEtAl14}. The relaxation rate can be calculated using the diagonal part of the linearized collision integral,
\begin{equation}
\begin{split}
\frac{1}{\tau(q_1)}=\frac{1}{2}\!\sum_{q_2,\qp{1},\qp{2}}\!\!W_{q_1,q_2}^{\qp{1},\qp{2}} \Bigl\{&N_B(\epsilon_{q_2})[1+N_B(\epsilon_{\qp{1}})+N_B(\epsilon_{\qp{2}})]
\\
&-N_B(\epsilon_{\qp{1}}) N_B(\epsilon_{\qp{2}})\Bigr\},
\end{split}
\end{equation}
where the transition probability is
\begin{equation}
W_{q_1,q_2}^{q_1^{\prime},q_2^{\prime}}=2\pi |\langle 0| b_{q_2^{\prime}}b_{q_1^{\prime}}\Ham_{\mathrm{nl}}b_{q_1}^{\dagger}b_{q_2}^{\dagger}|0\rangle|^2\delta(E_i-E_f),
\end{equation}
and $E_{i(f)}$ denotes the total energy of the initial (final) states. The modulus of the matrix element is given by
\begin{equation}
\begin{split}
|\langle 0 |b_{\qp{2}} b_{\qp{1}} \Ham_{\mathrm{nl}} b_{q_1}^{\dagger}b_{q_2}^{\dagger}|0\rangle|=\frac{\pi^2\Ej}{4\alpha^3N_x}&\sqrt{|q_1q_2 \qp{1}\qp{2}|}
\\
&\times\delta_{q_1+q_2,\qp{1}+\qp{2}}.
\end{split}
\label{Eq:matrix-element}
\end{equation}
A right moving plasmon with momentum $q_1>0$ can relax via this nonlinearity by the scattering off a left moving thermal plasmon with momentum $q_2<0$ (see Fig.~\ref{Fig:process}). According to the conservation laws, the momentum of the left moving particle
\begin{equation}
q_2=-\frac{3}{2}q_1\qp{1}\qp{2}l^2+\mathcal{O}(q_1^5 l^4)
\end{equation}
is much smaller than the momentum $q_1$.
With the help of the momentum conservation we can perform the sum over $\qp{2}$. The delta function related to the energy conservation can be written in the form
\begin{equation}
\delta(E_i-E_f)=\frac{2}{3u_0(q_1+q_2)|\qp{1,+}-\qp{1,-}|l^2}\delta(\qp{1}-\qp{1,+}), 
\end{equation}
where
\begin{equation}
\qp{1,\pm}\simeq \frac{q_1}{2}\pm\sqrt{\frac{q_1^2}{4}+\frac{2q_2}{3q_1 l^2}}.
\end{equation}
Requiring $\qp{1,\pm}$ to be real restricts the range of $q_2$ to the interval
\begin{equation}
q_{\ast}<q_2<0, \qquad q_{\ast}=-\frac{3}{8}q_1^3 l^2.
\end{equation}
In the continuum (long-chain) limit, the remaining summations over momenta transform into integrations,  $N_x^{-2}\sum_{q_2,\qp{1}}\to \int \diff q_2 \diff \qp{1}/(2\pi)^2$. Performing the integration over $\qp{1}$, we find
\begin{equation}
\begin{split}
\frac{1}{\tau(q_1)}=&\frac{\pi^3 \Ej^2 q_1}{96\alpha^6}\int_{q_{\ast}}^0\diff q_2 \frac{\qp{1,+}\qp{1,-}|q_2|}{u_0(q_1+q_2)|\qp{1,+}-\qp{1,-}|l^2}
\\
&\times \Bigl\{N_B(\epsilon_{q_2})[1+N_B(\epsilon_{\qp{1,+}})+N_B(\epsilon_{\qp{1,-}})]
\\
&\hspace{2.5cm}-N_B(\epsilon_{\qp{1,+}})N_B(\epsilon_{\qp{1,-}})\Bigr\}.
\end{split}
\label{Eq:final-integral}
\end{equation} 
The $q_2$-dependence in the denominator of the integrand can be neglected compared to $q_1$. We assume now that the energy of the particle with momentum $q_1$ is much larger than temperature but not too large such that $\beta u_0q_1^3l^2\ll 1$. In this case, the Bose function of the particle with momentum $q_2$ can be replaced by $1/\beta u_0 |q_2|$, and the Bose function related to the particle with momentum $q_{1,+}^{\prime}$ can be neglected. The Bose function related to the particle with momentum $q_{1,-}^{\prime}$ can be replaced by $1/\beta u_0\qp{1,-}$ for $-3q_1^2 l^2/2\beta u_0<q_2<0$ and neglected for $q_{\ast}<q_2<-3q_1^2 l^2/2\beta u_0$. The main contribution to the integral in Eq.~(\ref{Eq:final-integral}) originates from the latter range of $q_2$. After performing the integration, we find (under the assumptions $\beta u_0 q_1 \gg 1$, $q_1^2 l^2 \ll 1$, and $\beta u_0 q_1^3 l^2 \ll 1$) the following behavior of the relaxation rate:
\begin{equation}
\frac{1}{\tau(q_1)}\simeq \frac{\pi^3\Ej^2 Tq_1^4}{768\alpha^6u_0^2}=\frac{\pi^3}{768\alpha^5}\frac{\Ej}{E_0}Tq_1^4.
\label{Eq:relaxation-rate-not-too-large-q}
\end{equation}

Equation (\ref{Eq:relaxation-rate-not-too-large-q}) constitutes the main result of this section.  It predicts 
$\omega^4$ scaling of the relaxation rate of plasmons due to the ``gradient'' anharmonicity. 
The relaxation rate vanishes at low frequencies reflecting the irrelevant character of the gradient anharmonicities. 

Before closing this section let us discuss the universality of the result (\ref{Eq:relaxation-rate-not-too-large-q}) with respect to the particular form of the Hamiltonian $\Ham_{\rm nl}$ given by Eq.~(\ref{Eq:nonlinearity}). 
On phenomenological grounds various terms of the form $(\partial_x \phi)^n (\partial_x\theta)^m$ are allowed in the effective Hamiltonian.  For $n+m>4$ such terms are less relevant than the $(\partial_x \theta)^4$ term considered here and, thus, contribute less to the lifetime of plasmons. On the other hand a cubic-in-density interaction, $(\partial_x\phi)^3$,   is more relevant in terms of the scaling dimension. However, the energy and momentum conservations forbid the decay of a single plasmon into two particles. Correspondingly, a cubic non-linearity should be taken in the second order perturbation theory to produce a finite decay rate. The resulting process is again the one of Fig.~\ref{Fig:process} and leads to the same $\omega^4$ scaling of the relaxation rate\cite{Apostolov2013, Lin2013,  ProtopopovEtAl14}.

\subsection{Interplay of QPS and ``gradient'' anharmonicities}
\label{Sec:interplay}

In Secs.~\ref{Sec:phase_slips} and \ref{Sec:nonlinearities} we have analyzed the decay of plasmon excitations due to QPS and the ``gradient'' anharmonicities, respectively. While the relaxation rate due to ``gradient'' anharmonicities follows universal $\omega^4$ scaling, the QPS contribution is characterized by a non-universal exponent and reflects the SIT controlled by the value of $K_0$.
Let us now discuss  the interplay of the two relaxation channels. We assume for definiteness that frequencies of interest are larger than temperature. 
 
\begin{figure}
\centering
\includegraphics[scale=0.35]{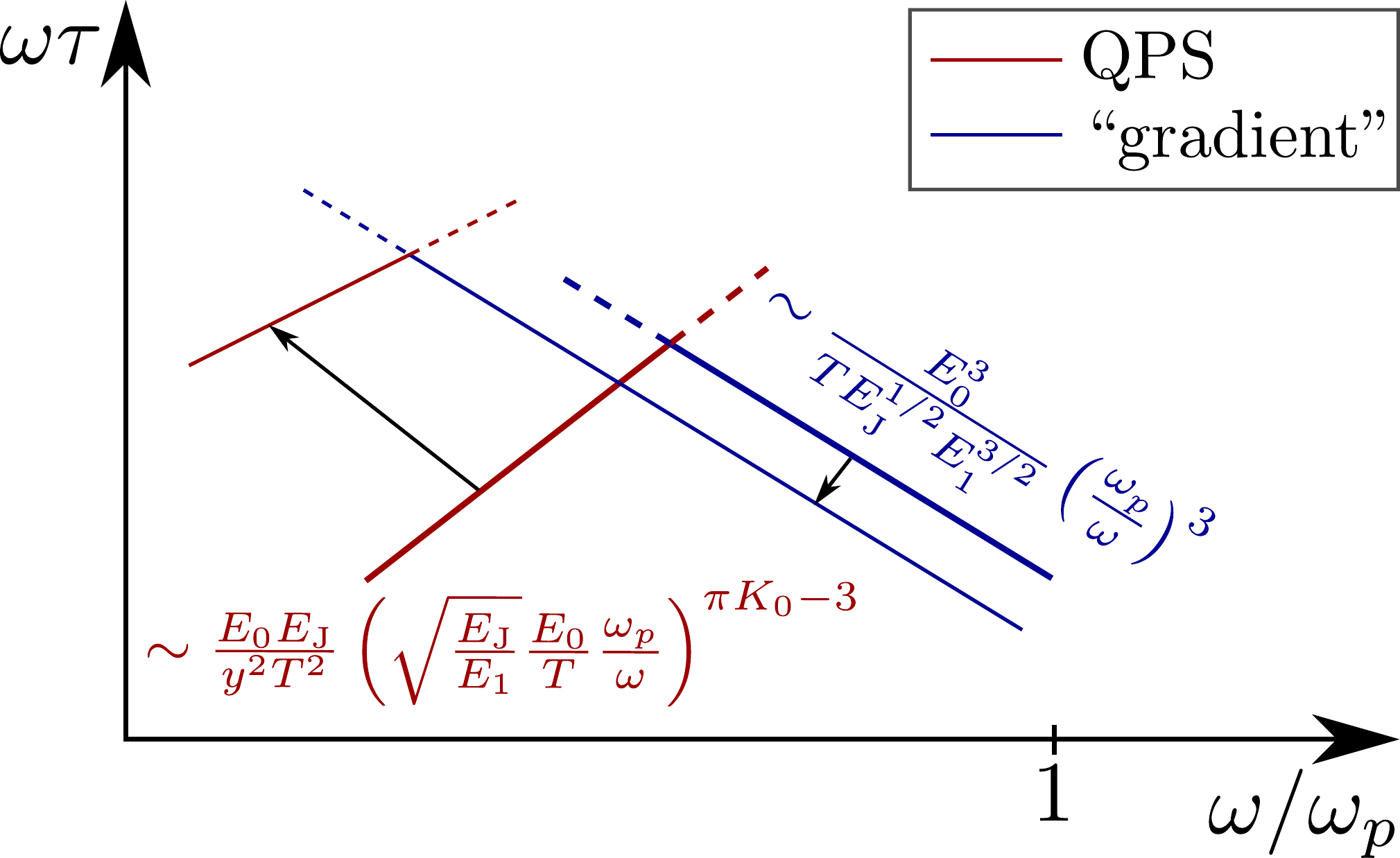}
\caption{Schematic behavior of the quality factor as a function of the rescaled frequency $\omega/\omega_p$  in a double-log scale in the insulating regime, $E_0 \gg \Ej$. The arrows indicate the change under an increase of $\Ej$ (thin lines correspond to a larger value of $\Ej$). The frequency scale of the crossover between the regime of dominant relaxation due to QPS (red lines) and that of dominant relaxation due to the nonlinearity (blue lines) is exponentially small in the parameter $\sqrt{\Ej/E_1}$. A further increase of $\Ej$ into the superconducting regime leads to a monotonic dependence of the quality factor (not shown in the figure). The scaling of the QPS and the ``gradient'' anharmonicity contributions indicated near the corresponding lines is based on Eqs.~\eqref{Eq:scaling-rate-clean} and \eqref{Eq:relaxation-rate-not-too-large-q}, respectively, with an assumption $\omega \gg T$. For the QPS contribution, the formula corresponds to the clean case. In the disordered case, the scaling of the QPS contribution can be inferred from Eqs.~\eqref{Eq:scaling-rate-strongly-disordered} and \eqref{Eq:inter}; this does not affect the qualitative appearance of the plot.}
\label{Fig:quality-factor}
\end{figure}

It is convenient to characterize the strength of the plasmon decay by a dimensionless parameter $\omega\tau$. This parameter is expected to be proportional to the quality factor studied in Ref.~\onlinecite{KuzminEtAl18}.  Deep in the superconducting regime, $K_0\gg1$, the relaxation of plasmons is always dominated by the ``gradient'' anharmonicities and the quality factor $\omega\tau$ scales as $\omega^{-3}$, see Eqs.~(\ref{Eq:scaling-rate-clean}), (\ref{Eq:scaling-rate-strongly-disordered}), (\ref{Eq:inter}) and (\ref{Eq:relaxation-rate-not-too-large-q}). Upon decreasing the Luttinger parameter $K_0$, the QPS start to be visible in the quality factor.  Specifically, the QPS dominate the low-frequency behavior of the quality factor under the condition $\pi K_0 < 6$ (respectively, $\pi K_0< 3$), yielding its $\omega^{3-\pi K_0}$ (respectively $\omega^{3-2\pi K_0}$) scaling in the cases of weak (respectively, strong) disorder. 
Furthermore, as a result of QPS, for  sufficiently small $K_0$ ($\pi K_0<3$ for weak and $\pi K_0<3/2$ for strong disorder), the quality factor goes down as frequency decreases. The resulting frequency dependence of $\omega\tau$ will then be non-monotonic with a maximum around a crossover frequency where the QPS set in, as  illustrated in Fig.~\ref{Fig:quality-factor}.

In the discussion of the overall frequency dependence of the quality factor, it is important to keep in mind the  exponential smallness of the fugacity $y$, Eq.~(\ref{Eq:y}). Due to this fact, the 
frequency below which QPS dominate over ``gradient'' non-linearities is exponentially small for $\Ej\gg E_1$. 
As a result, even deep in the insulating regime, $K_0\ll 1$,  the quality factor is dominated by the gradient non-linearities and thus  {\it grows} with lowering the frequency in a wide frequency range if $\Ej\gg E_1$. Only at exponentially small frequencies this ``superconducting'' behavior crosses over to the decrease of the quality factor reflecting the insulating character of the system in the infrared limit. 

Our results compare well with the experimental findings of Ref.~\onlinecite{KuzminEtAl18}, as we discuss in more detail in Sec.~\ref{Sec:Summary}.

\section{Summary and discussion\label{Sec:Summary}}

We have studied the decay of plasmonic waves in JJ chains. Motivated by a recent experiment\cite{KuzminEtAl18}, we have considered, besides
a  single one-dimensional chain, also a model of two capacitively coupled linear chains. It has been shown that in the parameter regime where the capacitance to the ground ($C_g$) can be neglected, the theory for the antisymmetric mode in the double chain can be mapped onto a theory for a single chain. This was possible because the symmetric mode acquired a fast velocity due to the strong Coulomb interaction. 

Two sources for the relaxation of plasma waves have been considered. First, the damping originating from the scattering generated by QPS leads to a relaxation rate that scales with frequency as a power law with a nonuniversal exponent that depends on the parameter $K_0=\sqrt{\Ej/E_0}$. The scaling behavior of the relaxation rate  related to QPS in different parameter regimes is summarized in Fig.~\ref{Fig:rate-QPS-processes}. Since the QPS amplitude is exponentially small in the parameter $\sqrt{\Ej/E_1}$, the rate is very sensitive to this parameter. The second mechanism for the relaxation of plasma waves is the interaction of them mediated by other nonlinear terms. As an example, we have considered the lowest-order nonlinearity coming from the Josephson potential. This term leads to a relaxation rate that scales as the fourth power of frequency. The vanishing of the relaxation rate at low frequencies reflects the irrelevance of this term in the renormalization group sense. Nevertheless, for a small phase-slip amplitude (fugacity), the contribution originating from this nonlinearity may be dominating in a wide range of frequencies. 

Comparing our findings to the experiment of Ref.~\onlinecite{KuzminEtAl18}, we find a very good qualitative agreement between our theory and experimental observations. All of the samples shown in Fig.~3b of Ref.~\onlinecite{KuzminEtAl18} are nominally  in the insulating regime. Specifically, values of the Luttinger constant $K_0$ that are extracted from the measured values of the impedance $Z$ (proportional to $1/K_0$) make one to expect the insulating behavior. 
However, the samples with a large ratio of $\Ej/E_1$ show an increase of the quality factor when lowering the frequency. This behavior suggests that the systems are in the superconducting regime. This apparent contradiction is resolved by noting that the crossover scale below which the QPS effects show up is exponentially small in the square root of $\Ej/E_1$.  As a result, the downturn of the quality factor indicating insulating behavior occurs below the lowest measured frequencies.
For devices with a lower value of both $K_0$ and $\Ej/E_1$, the authors  of Ref.~\onlinecite{KuzminEtAl18} observe a flat behavior at intermediate frequencies with a tendency to drop at lowest measured frequencies.  This behavior is qualitatively consistent with our prediction on the frequency dependence of the quality factor that is dominated by QPS in the insulating regime at low frequencies. For a more quantitative comparison, the extension of the experimental measurement method to lower frequencies and the investigation of the temperature dependence would be beneficial.

Let us discuss in more detail experimental observations on dependences of the quality factor on various input parameters. We consider first the more insulating chains. The authors of Ref.~\onlinecite{KuzminEtAl18} point out a  stronger sensitivity of the quality factor to the parameter $\Ej/E_1$ compared to the parameter $Z\propto 1/K_0$ for their weakest junctions (large $Z$ and low $\Ej/E_1$)---an observation that is  immediately understood within our theory.
In these devices, the parameter $K_0$ is very small such that the exponent for the power law of the phase-slip contribution to the quality factor is only slightly modified when changing $K_0$. Even relatively large changes of the order of $20\%$ (as in the experiment) have only a small effect, since the value of $K_0$ is still small and modifies the exponent only weakly. On the other hand, the fugacity of QPS depends exponentially on the square root of $\Ej/E_1$, which explains the observed strong dependence of the quality factor on this parameter. 
 
Further, we  compare the scaling predicted in our work to the experimental observations in low-impedance chains shown in Fig.~S~4 in the Supplementary Material of Ref.~\onlinecite{KuzminEtAl18}. All these samples are characterized by a large ratio of $\Ej$ over $E_1$ such that the QPS effects should be negligibly small in the range of measured frequencies. Indeed,  the curves show an increasing behavior when lowering the frequency. More specifically, the corresponding frequency scaling of the quality factor is consistent with the theoretical expectation $\omega^{-3}$ from the decay due to the nonlinearity. 
Discussing the dependence on other parameters, we notice that the charging energy $E_0$  experiences a particularly strong variation in the experiment (within a factor of $\sim 75$), while the variation of other device parameters is smaller. All experimental curves appear to collapse reasonably well when plotted as a function of the rescaled frequency $\omega/\omega_p$. On the other hand, our prediction shows a strong power-law dependence ($E_0^3$)  on the charging energy $E_0$. We speculate that a different kind of nonlinearity may be responsible for the explanation of this discrepancy. It might originate from some kind of nonlinear capacitances and result in a different prefactor in the frequency dependence of the quality factor that does not depend so strongly on the charging energy $E_0$. The identification and analysis of other types of nonlinearities constitutes an interesting prospect for future research.

Before closing this paper, we add two more comments on possible extensions of this work. First, we assumed that the Josephson and charging energy are constant for the whole chain. In principle, one can generalize the model by including spatial fluctuations of them. This will make the Luttinger-liquid constant $K_0$ randomly space dependent, $K_0\to K_0(x)$, and result in a possibility of elastic backscattering of plasmons that gets stronger with increasing frequency \cite{GramadaRaikh97,Fazio98}. In the experiment of Ref.~\onlinecite{KuzminEtAl18}, this disorder appears to be very weak, as can be inferred from regularly spaced resonances at higher frequencies. One can imagine, however, chains with a stronger $K_0(x)$-type disorder. An investigation of the combined effect of such a disorder and interaction on plasmon spectroscopy is an interesting prospect for future research. 

Second, our analysis of the width of the plasmonic resonances which relies on the golden rule assumes a continuous spectrum. This is justified if the obtained rate is larger than the the level spacings of final states to which a plasmon decays. In particular, for the gradient-anharmonicity decay, these are {\it three-particle} states: the final states for a decay of a  plasmon with momentum $q_1$ are characterized by three momenta $q_{1}^{\prime}$, $q_2$, and $q_2^{\prime}$, see Fig.~\ref{Fig:process}. The corresponding three-particle level spacing is much smaller than the single-particle level spacing in a long chain since it scales as $1/N_x^3$ with the length $N_x$. Thus, the analysis remains applicable despite the discrete \emph{single-particle} spectrum. The situation changes in shorter chains where one might be able to reach a regime in which the golden-rule rate is smaller than the three-particle level spacing. In this case, effects of localization in the Fock space may become important. For a related discussion in the context of electronic levels in quantum dots see Refs.~\onlinecite{Sivan94,SivanYmryAronov94,AltshulerGefenKamenevLevitov97,MirlinFyodorov97}.

While preparing this paper for publication, we learnt about a related unpublished work \cite{HouzetGlazmanUnpublished}. 

\textit{Note added in proof}. Recently, a preprint \onlinecite{WuSau18} appeared where a similar problem was addressed. The results of Ref.~\onlinecite{WuSau18} for plasmon decay due to QPS are consistent with our findings; gradient anharmonicities were not considered there.

\begin{acknowledgments}

We thank D.~Abanin, L.~Glazman, R.~Kuzmin, V.~E. Manucharyan, and A.~Shnirman for fruitful discussions. This work was supported by the Russian Science Foundation under Grant No.\ 14-42-00044 and  by the Deutsche Forschungsgemeinschaft. IP acknowledges support by the Swiss National Science Foundation.

 \end{acknowledgments}


\appendix
\section{Derivation of low-energy field theory}

\label{App:Field_theory}
This appendix is devoted to the derivation of the low-energy field theory for the antisymmetric mode of the double-chain system.   Our  starting point is the lattice Hamiltonian~\eqref{Eq:lattice-Hamiltonian-double-chain}.  We denote by $E_g$, $E_0$ and $E_1$ the charging energies associated with the capacitance $C_g$, $C_0$ and $C_1$, respectively, and $E_i=(2e)^2/C_i$.  

The basic idea is that in the limit of small capacitance $C_g$ the associated charging energy $E_g$ suppresses the charge fluctuations in the symmetric mode (at least at long scales) leaving us with the antiymmetric mode as the only dynamical degree of freedom. 
This  observation was previously employed in the literature to obtain the low-energy theory of the 
antisymmetric mode, see Refs.~\onlinecite{ChoiEtAl98b,ChoiEtAl00}. Here we generalize  the results  of Refs~ \onlinecite{ChoiEtAl98b,ChoiEtAl00} to the case when the Coulomb interaction is long-ranged ($C_1\gg C_0$) and charge disorder is present in the system.  We show that the effective theory takes the form of the sine-Gordon model, Eqs.~(\ref{Eq:S_clean}), (\ref{Eq:S_0}) and  (\ref{Eq:S_ps_disordered}) supplemented by a ``gradient'' non-linearity term, Eq.~(\ref{Eq:nonlinearity-action}). 

The posed goal can be achieved in two different ways. In Sec.~\ref{Ap:FiledTheory}, we present a semi-quantitative derivation of our results from the field-theory description of the symmetric and antisymmetric modes in the double chain. A more microscopic analysis of the initial lattice model (leading to the same results) is carried out in Secs.~\ref{App:Local} and \ref{App:LongRange} for the cases of short-range ($C_1=0$) and long-range ($C_1\gg C_0$) Coulomb interaction, respectively.

\subsection{Heuristic derivation from the continuum field theory}
\label{Ap:FiledTheory}

We start our discussion of the effective theory for the antisymmetric mode in the double chain  from a heuristic derivation based on the field-theory description of the lattice model (\ref{Eq:lattice-Hamiltonian-double-chain}). The latter is derived in full analogy to the case of a single JJ chain. To this end, we introduce two fields $\phi_\up$ and $\phi_\down$ related to the charge in the lower and upper chain via $\partial_x\phi_\sigma=-\pi{\cal N}_\sigma$, as well as their combinations 
\begin{equation}
\begin{pmatrix}
\phi_s
\\
\phi_a
\end{pmatrix}
=\frac12
\begin{pmatrix}
1& 1
\\
1 & -1
\end{pmatrix}
\begin{pmatrix}
\phi_{\up}
\\
\phi_{\down}
\end{pmatrix}.
\label{Eq:rotation}
\end{equation}
In terms of  these fields, the quadratic part of the action corresponding to the lattice model (\ref{Eq:lattice-Hamiltonian-double-chain}) reads 
\begin{equation}
\begin{split}
S_0=\frac{1}{\pi^2}&\int\frac{\diff q}{2\pi}\frac{\diff \omega}{2\pi}\Biggl\{\left[\frac{(2e)^2 q^2}{C_g+C_1 q^2}+\frac{\omega^2}{\Ej}\right]|\phi_s(\mbf{q})|^2
\\
&+\left[\frac{(2e)^2 q^2}{2C_0+C_g+C_1 q^2}+\frac{\omega^2}{\Ej}\right]|\phi_a(\mbf{q})|^2\Biggr\}.
\end{split}
\label{Eq:S_0_App}
\end{equation}

The QPS can be accounted for by 
\begin{equation}
S_{\mathrm{ps}}=yu_0\!\!\int\!\! \diff x \diff \tau \left\{\cos\left[2\phi_{\up}\!+\!{\cal Q}_{\up}(x)\right]+\cos\left[2\phi_{\down}+{\cal Q}_{\down}(x)\right]\right\},
\label{Sps}
\end{equation}
where 
\begin{equation}
\mathcal{Q}_{\sigma}(x)=2\pi \int_{-\infty}^{x}\diff x^{\prime} Q_{\sigma}(x^{\prime})
\end{equation}
and $Q_{\up(\down)}(x)$ is the random charge in the upper (lower) chain.
Note that  in Eq.~(\ref{Sps}) we consider  QPS as happening independently in the upper and lower chains. This is justified provided that $E_1\equiv(2e)^2/C_1 \ll E_g, E_0\equiv (2e)^2/C_0$. The fugacity $y$ is then exponentially small in the parameter $\sqrt{\Ej/E_1}$.

 In the long wave-length limit, $q\ll \sqrt{C_g/C_1}\ll \sqrt{C_0/C_1}$, the quadratic action (\ref{Eq:S_0_App}) reduces to the Luttinger-liquid form
\begin{equation}
S_0=\sum_{\rho=s,a}\frac{1}{2\pi^2u_{0,\rho}K_{0,\rho}}\int \diff x \diff \tau [u_{0,\rho}^2 (\partial_x \phi_{\rho})^2+(\partial_{\tau}\phi_{\rho})^2],
\end{equation}
with
\begin{equation}
\begin{split}
u_{0,s}&=\sqrt{\Ej E_g},  
\\
K_{0,s}&=\frac12\sqrt{\frac{\Ej}{E_g}},
\end{split}
\quad
\begin{split}
u_{0,a}&=\sqrt{\Ej E_0/2},
\\
 K_{0,a}&=\sqrt{\frac{\Ej}{2E_0}}.
\end{split}
\end{equation}

Let us now consider the perturbative expansion of the partition function $Z$ in the fugacity $y$. The lowest 
non-vanishing correction arises in the second order and reads
\begin{equation}
\begin{split}
\delta Z&=\frac{y^2u_0^2}{4}\int \diff^2 r_1 \diff ^2 r_2 \frac{1}{|\mbf{r}_1-\mbf{r}_2|^{2\pi K_{0,s}}}
\\
&\times\Bigl<\cos[2(\phi_a(\mbf{r}_1)-\phi_a(\mbf{r}_2))+\mathcal{Q}_{\up}(x_1)-\mathcal{Q}_{\up}(x_2)]
\\
&\hspace{0.4cm}+\cos[2(\phi_a(\mbf{r}_1)+\phi_a(\mbf{r}_2))+\mathcal{Q}_{\up}(x_1)-\mathcal{Q}_{\down}(x_2)]
\\
&\hspace{0.4cm}+\cos[2(\phi_a(\mbf{r}_1)+\phi_a(\mbf{r}_2))-\mathcal{Q}_{\down}(x_1)+\mathcal{Q}_{\up}(x_2)]
\\
&\hspace{0.4cm}+\cos[2(\phi_a(\mbf{r}_1)-\phi_a(\mbf{r}_2))-\mathcal{Q}_{\down}(x_1)+\mathcal{Q}_{\down}(x_2)]\Bigr>_{0,a}.
\end{split}
\label{Eq:DeltaZ1}
\end{equation}
Here, $\mbf{r}=(x, u_{0, a}\tau)$. 
In Eq.~(\ref{Eq:DeltaZ1}) we have performed explicit averaging over the symmetric mode $\phi_s$ but kept the correlation functions of $\phi_a$ in the unevaluated form.  
Introducing 
\begin{equation}
{\cal Q}_{s}(x)=\mathcal{Q}_{\up}(x)+\mathcal{Q}_{\down}(x)\,, \qquad 
{\cal Q}_{a}(x)=\mathcal{Q}_{\up}(x)-\mathcal{Q}_{\down}(x),
\end{equation}
we find
\begin{multline}
\delta Z=y^2u_0^2 \int \diff^2 r_1 \diff ^2 r_2 \frac{\cos[{\cal Q}_s(x_1)-{\cal Q}_s(x_2)]}{|\mbf{r}_1-\mbf{r}_2|^{2\pi K_{0,s}}}
\times\\
\left<\cos[2\phi_a(\mbf{r}_1)+{\cal Q}_{a}(x_1)]\cos[2\phi_a(\mbf{r}_2)+{\cal Q}_{a}(x_2)]\right>_{0,a}.
\label{Eq:correction_Z_App}
\end{multline}

Assuming that we are in the regime $K_{0,s}\ll 1$, we can approximate $|\mbf{r}_1-\mbf{r}_2|^{2\pi K_{0, s}}$ by unity. 
If the charge disorder is weak, we can further replace $\cos[{\cal Q}_s(x_1)-{\cal Q}_s(x_2)]$ by unity. 
In this case, the integrations over $\mbf{r}_1$ and $\mbf{r}_2$  decouple
and we observe that the correction (\ref{Eq:correction_Z_App}) can be viewed as resulting from the effective action [cf. Eq.~(\ref{Eq:S_0_App}); we take into account that $C_g\ll C_0$]
\begin{eqnarray}
S^{\rm eff}&=&S_0^{\rm eff}+S_{\mathrm{ps}}^{\mathrm{eff}},\\
S_0^{\rm eff}&=&
\frac{1}{\pi^2}\int\frac{\diff q}{2\pi}\frac{\diff \omega}{2\pi}
\left[\frac{(2e)^2 q^2}{2C_0+C_1 q^2}+\frac{\omega^2}{\Ej}\right]|\phi_a(\mbf{q})|^2,\nonumber\\
\label{Eq:S01}
\\
S_{\mathrm{ps}}^{\mathrm{eff}}&=&\sqrt{2} y u_0 \int \diff^2 r \cos[2\phi_a(\mbf{r})+\mathcal{Q}_a(x)],
\end{eqnarray}
which (up to a redefinition of the fugacity $y$ by an unimportant numerical factor) reproduces Eqs.~(\ref{Eq:S_clean}), (\ref{Eq:S_0}) and  (\ref{Eq:S_ps_disordered}) of  the main text with $\alpha=2$. 

If the charge disorder is strong, we expand the cosine in Eq.~\eqref{Eq:correction_Z_App},
\begin{equation}
\begin{split}
\cos[{\cal Q}_s(x_1)-{\cal Q}_s(x_2)]&=\cos[{\cal Q}_s(x_1)]\cos[{\cal Q}_s(x_2)]
\\
&\,\,+\sin[{\cal Q}_s(x_1)]\sin[{\cal Q}_s(x_2)].
\end{split}
\label{Eq:expansion-cosine}
\end{equation} 
Both terms in Eq.~\eqref{Eq:expansion-cosine}, when substituted into Eq.~\eqref{Eq:correction_Z_App},  give equivalent contributions, if the disorder ${\cal Q}_s$ is strong. In the opposite limit of weak disorder (small ${\cal Q}_s$), the second term would be much smaller than the first one. Thus, keeping only the first term will always yield a correct result, up to a coefficient of order unity. Proceeding in this way, we again find an effective action for QPS that is of first order in $y$ [cf. discussion of the weakly disordered case],
\begin{equation}
\label{S-ps-random-fugacity}
S_{\mathrm{ps}}^{\mathrm{eff}}=\sqrt{2}y u_0 \int \diff^2 r \cos[\mathcal{Q}_s(x)]\cos[2\phi_a(\mbf{r})+\mathcal{Q}_a(x)].
\end{equation}
For strong charge disorder we find, besides the random phase, also a random amplitude of the QPS action. As shown in Ref.~\onlinecite{BardEtAl17}, the QPS action without a random amplitude, Eq.~\eqref{Eq:S_ps_disordered}, automatically generates a QPS term with a random amplitude if the charge disorder is strong. The phase-slip action Eq.~\eqref{Eq:S_ps_disordered} hence adequately describes the effects of QPS on the antisymmetric mode in the double chain in the disordered case.

Let us now discuss the ``gradient'' anharmonicity correction to the effective action of the antisymmetric mode. Taking into account the ``gradient'' anharmonicity arising from the quartic expansion of the Josephson coupling in each of the two chains one finds 
\begin{equation}
S_{\mathrm{nl}}=\!\frac{-1}{12\pi^4 \Ej^3}\!\!\int \!\!\diff x\! \left[(\partial_{\tau} \phi_a)^4\!+\!(\partial_{\tau} \phi_s)^4\!+\!6(\partial_{\tau} \phi_s)^2(\partial_{\tau} \phi_a)^2\right].
\label{Eq:nonlinear-both-modes}
\end{equation}
We now average (\ref{Eq:nonlinear-both-modes}) over fluctuations of $\phi_s$. Omitting a trivial constant term arising from the first term in Eq.~(\ref{Eq:nonlinear-both-modes}) and a renormalization of the Josephson energy in Eq.~(\ref{Eq:S01}) by a numerical factor arising from the last term we get
\begin{equation}
S_{\rm nl}^{\rm eff}=-\frac{1}{12\pi^4 \Ej^3}\int \diff x \diff \tau\,(\partial_{\tau} \phi_a)^4,
\end{equation}
and reproduce Eq.~(\ref{Eq:nonlinearity-action}) with $\alpha=2$.

\subsection{Elimination of symmetric mode at the level of the lattice model: the case of local Coulomb interaction} 
\label{App:Local}
In this appendix we  assume local Coulomb interaction ($C_1=0$) and derive the effective theory of the antisymmetric mode by integrating out the symmetric mode  directly in the lattice model (\ref{Eq:lattice-Hamiltonian-double-chain}). We closely follow here the derivation of the effective theory for a single chain outlined in appendix A of Ref.~\onlinecite{BardEtAl17}. The generalization of this derivation to the case $C_1\gg C_0$ will be presented in Sec.~\ref{App:LongRange}.

We start by constructing the path-integral representation of the partition function for the system. To this end, we discretize the (imaginary) time $\tau \in [0,\beta)$ in $N_{\tau}$ steps with spacing $\Delta \tau$ (the precise value will be discussed later). For concreteness we assume periodic boundary conditions along the chains, with $N_x$ grains in each chain. In the following, $n$ and $i$ are the indices of the lattice point in $\tau$ and  $x$ directions, respectively, and $\sigma=\up,\down$ discriminates between the two chains. At each vertex of the space-time lattice ($n$,$i$,$\sigma$), a resolution of unity of the form
\begin{equation}
\begin{split}
\mathds{1}=\sum_{\mathcal{N}_{\up},\mathcal{N}_{\down}}\int_{0}^{2\pi}\frac{\diff \theta_{\up}}{2\pi}\int_{0}^{2\pi}\frac{\diff \theta_{\down}}{2\pi}&\left|\mathcal{N}_{\up},\mathcal{N}_{\down}\right\rangle\left\langle\theta_{\up},\theta_{\down}\right|
\\
&\times \e^{-i \theta_{\up}\mathcal{N}_{\up}}\,\e^{-i \theta_{\down}\mathcal{N}_{\down}}
\end{split}
\end{equation}
is inserted. This results in the action
\begin{equation}
\begin{split}
&S=-i\sum_{n,i,\sigma}\mathcal{N}_{i,\sigma}^n(\partial_{\tau}\theta)^n_{i,\sigma}+\Ej \Delta\tau \sum_{n,i,\sigma} (1-\cos[(\partial_x \theta)_{i,\sigma}^n])
\\
&\!+\frac{(2e)^2\Delta\tau}{2}\!\!\sum_{n,i,\sigma,\sigma^{\prime}}\!(C^{-1})_{\sigma,\sigma^{\prime}}\left(\mathcal{N}_{i,\sigma}^n-Q_{i, \sigma}\right)\left(\mathcal{N}_{i,\sigma^{\prime}}^n-Q_{i, \sigma^\prime}\right),
\end{split}
\end{equation}
where we have introduced the lattice derivatives
\begin{equation}
(\partial_x\theta)_{i,\sigma}^n=\theta_{i+1,\sigma}^n-\theta_{i,\sigma}^n
\,\, \mathrm{and} \,\,
(\partial_{\tau}\theta)_{i,\sigma}^n=\theta_{i,\sigma}^{n+1}-\theta_{i,\sigma}^n;
\end{equation}
by $Q_{i, \sigma}$ we denote the stray charges and 
the inverse capacitance matrix in the local case reads
\begin{equation}
C^{-1}=\frac{1}{C_g(C_g+2C_0)}
\begin{pmatrix}
C_g+C_0 & C_0
\\
C_0 & C_g+C_0
\end{pmatrix}.
\end{equation}

To perform the summation over the charge variables ${\cal N}^n_{i, \sigma}$, it is convenient to introduce the symmetric and antisymmetric combinations of charges and phases
\begin{align}
{\cal N}^n_{i, s}&=\frac{{\cal N}^n_{i, \up}+ {\cal N}^n_{i, \down}}{2}, 
&{\cal N}^n_{i, a}&=\frac{{\cal N}^n_{i, \up}- {\cal N}^n_{i, \down}}{2},
\label{Eq:Nsa}\\
{Q}_{i, s}&=\frac{Q_{i, \up}+ Q_{i, \down}}{2}, 
&Q_{i, a}&=\frac{Q_{i, \up}- Q_{i, \down}}{2} \label{Eq:Qrot},\\
\theta_{i,s}^n&=\frac{\theta_{i,\up}^n+\theta_{i,\down}^n}{2}, &\theta_{i,a}^n&=\theta_{i,\up}^n-\theta_{i,\down}^n.
\label{Eq:thetaRot}
\end{align}

According to (\ref{Eq:Nsa}), the charges  ${\cal N}^n_{i, s}$ and  ${\cal N}^n_{i, a}$ are either  both integer or both half-integer. Note also the absence of $1/2$ in the definition of $\theta_{i, a}^n$.

The partition function reads now
\begin{equation}
Z=\sum_{\{{\cal N}_{i,s}^n,{\cal N}_{i,a}^n\}}\int_{0}^{2\pi} \mathcal{D}\theta_{\up}\mathcal{D}\theta_{\down} \e^{-\sum_{i,n}S_i^n},
\label{Eq:partition_function_new_variables}
\end{equation}
with 
\begin{equation}
\begin{split}
S_i^n=&-2 i {\cal N}_{i,s}^n(\partial_{\tau}\theta)_{i,s}^n
-i {\cal N}_{i,a}^n(\partial_{\tau}\theta)_{i,a}^n
\\
&+(2e)^2\Delta \tau \Bigl[\frac{({\cal N}_{i,s}^n-Q_{i, s})^2}{C_g}+\frac{({\cal N}_{i,a}^n-Q_{i, a})^2}{2C_0+C_g}\Bigr]
\\
&+\Ej \Delta\tau \sum_{\sigma}(1-\cos[(\partial_x\theta)_{i,\sigma}^n]).
\end{split}
\label{Eq:S1}
\end{equation}
We observe that  in the limit of a small capacitance $C_g$, $(2e)^2/C_g\gg \Ej, E_0$, 
the dynamics of the charges    ${\cal N}_{i, s}$ is frozen out and their  values are pinned to the background charges $Q_{i, s}$\footnote{In the case of strong charge disorder we neglect here the rare sites in the chain where $2Q_{i, s}$ is half integer.}
\begin{equation}
{\cal N}_{i, s}^{n}=\frac12 \lfloor{2 Q_{i, s}}  \rfloor
\end{equation}
where $\lfloor{\cdot}\rfloor$ stands for the integer part. The first term in Eq.~(\ref{Eq:S1}) is then a total derivative and can be dropped due to periodic boundary conditions in the imaginary time. Moreover, it is easy to see that, upon the proper redefinition  of the stray charges $Q_{i, a}$, one can regard the summation over ${\cal N}_{i, a}^n$ in the partition function as running over integers irrespective of  the (integer or half integer) value of ${\cal N}_{i, s}$. We thus conclude that, with the charges ${\cal N}_{i, s}$ being frozen out, the 
dynamics of the system is governed by the action
\begin{equation}
\begin{split}
S=\sum_{n,i}\Bigl\{&-i {\cal N}_{i,a}^n(\partial_{\tau}\theta)_{i,a}^n+(2e)^2 \Delta \tau \frac{({\cal N}_{i,a}^n-Q_{i, a})^2}{C_g+2C_0}
\\
&+2\Ej \Delta\tau \left(1-\cos[(\partial_x\theta)_{i,s}^n]\cos[(\partial_x\theta)_{i,a}^n/2]\right)\Bigr\}.
\end{split}
\label{Eq:S2}
\end{equation}

The last step one needs to perform in order to derive from Eq.~(\ref{Eq:S2}) the effective action for the antisymmetric mode is the integration over the phases $\theta_{i, s}^n$. To this end, we assume open boundary conditions in the space direction and introduce new integration variables 
\begin{equation}
\tilde{\theta_{i}^n}=\theta_{i, s}^n-\theta_{i-1, s}^n, \qquad i\geq 2.
\end{equation}
The relevant factor in the partition function takes then the form 
\begin{equation}
\begin{split}
&\prod_{i=1}^{N_x-1} \prod_{n=1}^{N_{\tau}}\Bigl(\int_0^{2\pi} \diff \tilde{\theta}_{i+1,s}^n
\\
&\times \exp\left\{-2\Ej \Delta\tau (1-\cos[(\partial_x\theta)_{i,a}^n/2]\cos[\tilde{\theta}_{i+1,s}^n])\right\}\Bigr)\\
& \propto \exp\left\{-\Delta \tau \sum_{i=1}^{N_x-1} \sum_{n=1}^{N_{\tau}} g
\left[(\partial_x\theta)_{i,a}^n\right]\right\}.
\end{split}
\end{equation}
Here we have dropped an irrelevant normalization factor, and the function $g(\gamma)$ can be expressed in terms of the modified Bessel function $\mathrm{I}_0(\gamma)$:
\begin{equation}
g(\gamma)=-\frac{1}{\Delta\tau}\ln \mathrm{I}_0\left(2\Ej\Delta\tau\cos \frac\gamma2\right). 
\label{Eq:g}
\end{equation}

The function $g(\gamma)$ is $2\pi$ periodic in its argument. Thus, we can regard the effective action 
of the antisymmetric mode,
\begin{multline}
S=\sum_{n,i}\Bigl\{-i {\cal N}_{i,a}^n(\partial_{\tau}\theta)_{i,a}^n+(2e)^2 \Delta \tau \frac{({\cal N}_{i,a}^n-Q_{i, a})^2}{C_g+2C_0}\\+ \Delta\tau \, g[(\partial_x\theta)_{i,a}^n]\Bigr\},
\end{multline}
as describing a chain of JJs with the effective Josephson coupling given by $g(\partial_x \theta)$ and proceed  in close analogy with Ref.~\onlinecite{BardEtAl17}.  We develop the theory 
starting from the superconducting ground state. 
 As we will see later, this means that we are in the limit $\Ej \Delta \tau \gg 1$. In this limit, the main contribution comes from the region close to $\partial_x\theta_a=0$ (mod $2\pi$). Thus, we can 
 employ the Villain approximation that reads 
\begin{equation}
\exp\left[ -\Delta\tau \, g(\partial_x \theta_a)\right]\propto \sum_{h}\e^{-\frac{\Ej\Delta\tau}{4}(\partial_x\theta_a-2\pi h)^2}.
\end{equation}
Fixing the time step $\Delta\tau$ to
\begin{equation}
\Delta \tau=\sqrt{\frac{2}{\Ej E_0}}=\sqrt{\frac{2C_0}{(2e)^2\Ej}},
\end{equation}
and following the derivation of the sine-Gordon theory discussed in Ref.~\onlinecite{BardEtAl17}, we find (skipping the index ``a'')
\begin{equation}
\begin{split}
S=\frac{1}{2\pi^2 K_0}\int \diff x\diff \tau& [u_0^2(\partial_x\phi)^2+(\partial_{\tau}\phi)^2]
\\
&+yu_0 \int \diff x \diff \tau \cos[2\phi(x,\tau)+{\cal Q}_a],
\end{split}
\label{S_local_App}
\end{equation}
where 
\begin{equation}
 K_0=\sqrt{\frac{\Ej}{2E_0}}\,, \qquad u_0=\sqrt{\Ej E_0/2}.
\label{Eq:definition_K}
\end{equation}

Equation (\ref{S_local_App}) is equivalent to Eqs.~(\ref{Eq:S_clean}), (\ref{Eq:S_0}) and  (\ref{Eq:S_ps_disordered})  in the limit $\Lambda=\infty$. To complete our analysis we thus only need  to extract the ``gradient'' anharmonicity term. It arises from the fourth order expansion of the effective Josephson coupling (\ref{Eq:g}) and reads
\begin{equation}
\Ham_{\mathrm{nl}}=-\frac{\Ej}{192}\int \diff x \left(\partial_x\theta_a\right)^4.
\label{Eq:nonlinearity_App}
\end{equation}
This result coincides with the ``gradient'' anharmonicity term stated in the main text, Eq.~(\ref{Eq:nonlinearity}), with $\alpha=2$.

\subsection{Elimination of symmetric mode at the level of the lattice model: the case of long-range Coulomb interaction}

\label{App:LongRange}
Let us now discuss the derivation of the effective theory in the case of the long-range Coulomb interaction, $C_0\ll C_1$. Throughout this section we take the limit of $C_g=0$. 

It is convenient to represent the partition function as a path integral over the phases $\theta_{i, \sigma}(\tau)$
\begin{equation}
Z=\int \prod_{i,\sigma} {\cal D}\theta_{i, \sigma}(\tau) e^{-S}
\label{Eq:ZTheta}
\end{equation}
with the action
\begin{multline}
S=\int \diff\tau \left\{\sum_{i, \sigma}\left[\frac{[(\partial_x\dot{\theta})_{i, \sigma}]^2}{2E_1}
-\Ej
\cos\left[(\partial_x \theta)_{i, \sigma}\right]\right.\right.\\+\left.i\dot{\theta}_{i, \sigma}Q_{i, \sigma}\Biggr]+\sum_{i}\frac{\left(\dot{\theta}_{i,\up}-\dot{\theta}_{i, \down}\right)^2}{2E_0}.
\right\}
\label{Eq:STheta}
\end{multline}
The action Eq. (\ref{Eq:STheta}) is equivalent to the Hamiltonian (\ref{Eq:lattice-Hamiltonian-double-chain}) in the limit $C_g=0$. The first term in the second line describes the effect  of  random stray charges.
The quantization of the grain charges ${\cal N}_{i, \sigma}$ is reflected in the boundary condition along the imaginary time
\begin{equation}
\theta_{i, \sigma}(\beta)=\theta_{i, \sigma}(0)+2\pi n_{i, \sigma} \ ,
\end{equation}
 where $\beta$ is the inverse temperature and $n_{i, \sigma}$ are integers. 
 
 In the considered limit of $C_g=0$ the dependence of the action on the symmetric combination of phases, $\theta_{i, s}\equiv (\theta_{i, \up}+\theta_{i, \down})/2$  is through its spatial gradient only. 
 We thus introduce  
 \begin{equation}
 \Theta_{i,s}=\frac{(\partial_x \theta)_{i,\up}+\partial_x(\theta)_{i,\down}}{2}, \qquad \theta_{i,a}=\theta_{i,\up}-\theta_{i,\down}.
 \end{equation}
 as new integration variables and find
 \begin{equation}
 \begin{split}
 S&=\int \diff\tau \sum_{i}\Biggl\{
 \frac{\dot{\Theta}_{is}^2}{E_1}
+ \frac{\left[(\partial_x\dot{\theta})_{i, a}\right]^2}{4E_1}
+2i\dot{\Theta}_{i, s}{\cal Q}_{i, s}
\\
&+i\dot{\theta}_{i, a} Q_{i, a}-2\Ej \cos\left[\Theta_{i, s}\right]\cos\left[\frac{(\partial_x \theta)_{i, a}}{2}\right]+\frac{\dot{\theta}_{i,a}^2}{2E_0}
\Biggr\}.
\end{split}
\label{Eq:STheta1}
 \end{equation}
 Here 
 \begin{equation}
 {\cal Q}_{i, s}=\sum_{j<i}Q_{j, s}
 \end{equation}
 and the symmetric and antisymmetric combinations of the stray charges, $Q_{i, s}$ and $Q_{i, a}$, are defined according to Eq.~(\ref{Eq:Qrot}). 
The boundary conditions in the time direction are given by
\begin{eqnarray}
\theta_{i, a}(\beta)&=&\theta_{i, a}(0)+2\pi n_{i, a} \ ,\\
\Theta_{i, s}(\beta)&=&\Theta_{i, s}(0)+2\pi n_{i, s}+\pi \delta_{i} \ ,
\end{eqnarray}
where $n_{i, s (a)}$ are integer numbers and
\begin{equation}
\delta_{i}=(n_{i+1, a}-n_{i, a})\mod 2.
\end{equation} 

We can now formally perform the functional integration over the symmetric mode. 
Indeed, the integrations at different spatial points decouple. It is then easy to see that the result of the integration over $\Theta_{i, s}(\tau)$ can be expressed as
\begin{multline}
\int {\cal D} \Theta_{i, s}(\tau) \exp\left\{-
\int \diff\tau \left[\frac{\dot{\Theta}_{is}^2}{E_1} +2i\dot{\Theta}_{i, s}{\cal Q}_{i, s} \right.\right.\\\left.\left.- 2\Ej
\cos\left[\Theta_{i, s}\right]\cos\left[\frac{(\partial_x \theta)_{i, a}}{2}\right]\right]
\right\}\\=\Tr U(\beta)\equiv e^{-\delta S[\partial_x \theta_{i, a}(\tau)]},
\label{Eq:deltaS}
\end{multline} 
where $U(\tau)$ is the (imaginary-time) evolution operator defined by
\begin{equation}
\frac{\diff U}{\diff \tau}=-{\cal H}[\theta_{i, a}(\tau)-\theta_{i+1, a}(\tau)]U(\tau),
\label{Eq:U}
\end{equation}
with the time dependent Hamiltonian
\begin{multline}
\mathcal{H}=E_1 \left({\cal N}-2{\cal Q}_{i, s}-\frac{\delta_i}{2}\right)^2\\-2\Ej\cos\left(\Theta +\pi \delta_i\frac{\tau}{\beta}\right)\cos\frac{\theta_{i+1, a}(\tau)-\theta_{i, a}(\tau)}{2}.
\label{Eq:HTheta}
\end{multline}
Here, ${\cal N}$ is the (integer-valued) momentum canonically conjugate to the coordinate $\Theta$.

The contribution  $\delta S[\partial_x \theta_{i, a}(\tau)]$ to the action of the antisymmetric mode defined by Eqs. (\ref{Eq:deltaS}), (\ref{Eq:U}) and  (\ref{Eq:HTheta}) is generally a complicated functional of the phase difference $\partial_x \theta_{i, a}(\tau)$. We are mainly interested, however, in the low-frequency modes of the field $\theta_{i, a}$ (with frequencies much less than the plasma frequency $\sqrt{E_1 \Ej}$). The adiabatic approximation can then be used for the computation of the evolution operator (\ref{Eq:U}). Moreover, for  $E_1\ll \Ej$ and low temperature, the dynamics of $\Theta$ can be determined just by the minimization of the potential energy in the Hamiltonian (\ref{Eq:HTheta}). This leads to
\begin{equation}
\delta S=-2\Ej\int \diff\tau \left|\cos\frac{(\partial_x \theta)_{i, a}}{2} \right| .
\label{Eq:DeltaS}
\end{equation}

Equations (\ref{Eq:STheta1}) and (\ref{Eq:DeltaS}) give rise to the effective action for the antisymmetric mode
\begin{multline}
S= \int \diff\tau \sum_{i}\left\{
 \frac{\left[(\partial_x\dot{\theta})_{i, a}\right]^2}{4E_1}
-2\Ej
\left|\cos\left[\frac{(\partial_x \theta)_{i, a}}{2}\right]\right|
\right.\\\left. +\dot{\theta}_{i, a} Q_{i, a}+\frac{\dot{\theta}_{i,a}^2}{2E_0}
\right\}.
\label{Eq:SThetaA}
\end{multline}
The subsequent derivation of the effective sine-Gordon theory proceeds then along the lines of Ref. \onlinecite{BardEtAl17} and leads to Eqs. (\ref{Eq:S_clean}), (\ref{Eq:S_0}) and  (\ref{Eq:S_ps_disordered}) with $\alpha=2$. The fourth order expansion of the Josephson coupling in 
(\ref{Eq:SThetaA}) gives rise to the ``gradient'' anharmonicity, Eq. (\ref{Eq:nonlinearity}).  

Before closing this section, let us comment on the relation between the presented derivation and the field-theoretic derivation discussed in Sec.~\ref{Ap:FiledTheory} of the appendix. Both derivations lead to the effective sine-Gordon model for the antisymmetric mode. It was found in Sec.~\ref{Ap:FiledTheory} that the corresponding fugacity $y$ fluctuates in space, see Eq.(\ref{S-ps-random-fugacity}). Such fluctuations are not seen in Eq. (\ref{Eq:SThetaA}). We anticipate that a more accurate treatment of QPS based on Eqs. (\ref{Eq:U}) and (\ref{Eq:HTheta}) will produce a fugacity $y$ that depends on the configuration of the stray charges $Q_{i, s}$ and fluctuates in space.   Furthermore,  as shown in Ref.~\onlinecite{BardEtAl17}, the random amplitude of the QPS term is generated under the renormalization-group transformation. The results of both derivations are therefore equivalent.


%

\end{document}